\documentclass[twocolumn]{aastex631}
\usepackage{multirow}
\usepackage{subfigure}
\usepackage{amsmath}
\usepackage{tabularx}
\usepackage{todonotes}

\begin{document}

\title{Probing the Extended Atmospheres of AGB Stars: \\I. Synthetic imaging of 1D hydrodynamical models at radio and (sub-)millimeter wavelengths}

\author[0000-0002-6901-4086]{Behzad Bojnordi Arbab}
\affiliation{Department of Space, Earth and Environment, Chalmers University of Technology}

\author[0000-0002-2700-9916]{Wouter Vlemmings}
\affiliation{Department of Space, Earth and Environment, Chalmers University of Technology}

\author[0000-0002-5680-9525]{Theo Khouri}
\affiliation{Department of Space, Earth and Environment, Chalmers University of Technology}

\author[0000-0003-2356-643X]{Susanne Höfner}
\affiliation{Theoretical Astrophysics, Department of Physics and Astronomy, Uppsala University, Box 516, 75120 Uppsala, Sweden}

\begin{abstract}

We investigate the observable characteristics of the extended atmospheres of AGB stars across a wide range of radio and (sub-)mm wavelengths using state-of-the-art 1D dynamical atmosphere and wind models over one pulsation period. We also study the relationships between the observable features and model properties. We further study practical distance ranges for observable sources assuming the capabilities of current and upcoming observatories. We present time-variable, frequency-dependent profiles of pulsating AGB stars’ atmospheres, illustrating observable features in resolved and unresolved observations, including disc brightness temperature, photosphere radius, and resolved and unresolved spectral indices. Notably, temporal variations in disc brightness temperature closely mirror the temperature variability of the stellar atmosphere. We find that while the photospheric radius decreases due to gas dilution in the layers between consecutive shocks, the increase in the observed stellar radius reflects shock propagation through the atmosphere during the expansion phase, providing a direct measurement method for the shock velocity. Furthermore, our models indicate that enhanced gas temperatures after the passage of a strong shock might be observable in the high-frequency ALMA bands as a decrease in the brightness temperature with increasing frequency. We demonstrate that synthetic observations based on state-of-the-art dynamical atmosphere and wind models are necessary for proper interpretations of current (ALMA and VLA) and future (SKA and ngVLA) observations and that multi-wavelength observations of AGB stars are crucial for empirical studies of their extended atmospheres.

\end{abstract}


\section{Introduction} 
\label{section:Introduction}
    Stars with main-sequence masses ranging from 1 to 8 solar masses undergo significant mass loss during the asymptotic giant branch (AGB) phase. This mass loss process profoundly influences the chemistry of the interstellar medium (ISM) and plays a pivotal role in shaping the subsequent evolutionary paths of stars \citep{Marigo2020, Karakas2022}. The mechanism behind this mass loss is believed to be the outflow of gas driven by dust particles accelerated outwards by radiation pressure. 
    These dust grains form from molecules in the cool and dense regions of extended atmospheres, created by stellar pulsations, convective motions, and shock waves. For a recent review on AGB star mass loss, see \citet{Hofner2018}.

    Observations conducted in the infrared, (sub)millimeter, and radio wavelengths are crucial for studying the properties and dynamics of the extended atmospheres of evolved stars. \citet[hereafter RM97]{Reid1997} partially resolved the continuum emissions of a radio atmosphere at centimeter wavelengths using the Very Large Array (VLA), revealing that AGB stars appear larger at these long wavelengths due to the presence of free-free opacity in their extended atmospheres.
    The Atacama Large Millimeter/submillimeter Array (ALMA) and VLA, with their exceptional sensitivity and high spatial resolution, have facilitated the mapping of molecules and dust distribution in the outer envelopes of AGB stars. 
    These observations have unveiled intricate structures and provided insights into the mechanisms driving mass loss \citep[e.g.][]{Matthews2015, Khouri2016, Khouri2024A&A}. 
    \citet[hereafter V19]{Vlemmings2019} reported observations of AGB stars with ALMA, which resolved the stellar continuum at mm wavelengths. They modeled the observations using power laws for the density, temperature, and ionization radial profiles and concluded that layers of enhanced opacity are necessary to explain the data. Furthermore, \hyperlink{V19}{V19} observations of R Leo at ALMA bands 4 and 6 within 2 weeks during 0.41-0.46 phase revealed the outward motion of mm-wavelength surface at the speed of $10.6 \pm 1.4~{\rm km~s^{-1}}$.
    At shorter wavelengths, the Very Large Telescope Interferometer (VLTI) has enabled precise measurements of angular sizes for AGB stars and their surrounding molecular and dust shells, aiding in the characterization of their geometry and kinematics \citep[e.g.][]{Wittkowski2017}.
    
    Resolved observations of the extended atmospheres of evolved stars are limited to only a few objects, and temporal variations in radio and mm frequencies have not been thoroughly studied.
    In a most recent study, \citet[]{Vlemmings2024} reported a series of resolved observations of R Dor with ALMA in bands 6 and 7, and presented small-scale features on the surface of the star. Using the temporal variations of the stellar surface, they measured the radial velocity profile of large convection cells and determined the timescale for the surface structures to be approximately $33$~days.
    
    Dynamical models have addressed critical issues related to mass-loss processes in AGB stars, such as convection, pulsation, and dust growth. In most cases, these models assumed spherically symmetric structures \citep[e.g.][]{Winters2000, Jeong2003, Hofner2016, Hofner2022}, but more recently also full 3D geometry has been taken into account \citep[e.g.][]{hofner2019, Freytag2023}.
    The predictions from these models for the innermost regions of the circumstellar envelope can be directly compared to high-resolution observations, providing a crucial test of our theoretical understanding.

    Given the capabilities of recent and upcoming observatories, it will soon be possible to study the resolved extended atmospheres across a uniquely wide wavelength range. This will allow us to characterize gas with a broader range of densities and temperatures, measure structure sizes and radial velocities, and compare the observations directly to state-of-the-art (multi-dimensional) models. 
    In this paper, we analyze synthetic images of extended atmospheres using state-of-the-art radiation-hydrodynamical atmosphere and wind models \citep[DARWIN,][]{Hofner2022}, showing the observable characteristics and variations across a wide range of frequencies. 
    While observable properties of these models in the visual and infrared regime have been discussed before, this is the first time that synthetic data for radio and mm wavelengths is presented.
    Leveraging these models and the capabilities of current and future facilities, we highlight the potential to comprehensively characterize the extended atmospheres of evolved stars through resolved and unresolved observations. We outline our models and discuss radiative transfer in Section \ref{section:Methods}, present the results of our model analysis and imaging in Section \ref{section:Results}, assess the capabilities of current and future observational facilities in radio and mm frequencies in Section \ref{section:Discussion}, and conclude with insights into the potential for future observations in Section \ref{section:Conclusion}.

\section{Methods}
\label{section:Methods}

    \begin{figure*}
        \centering
        \includegraphics[width=1\linewidth]{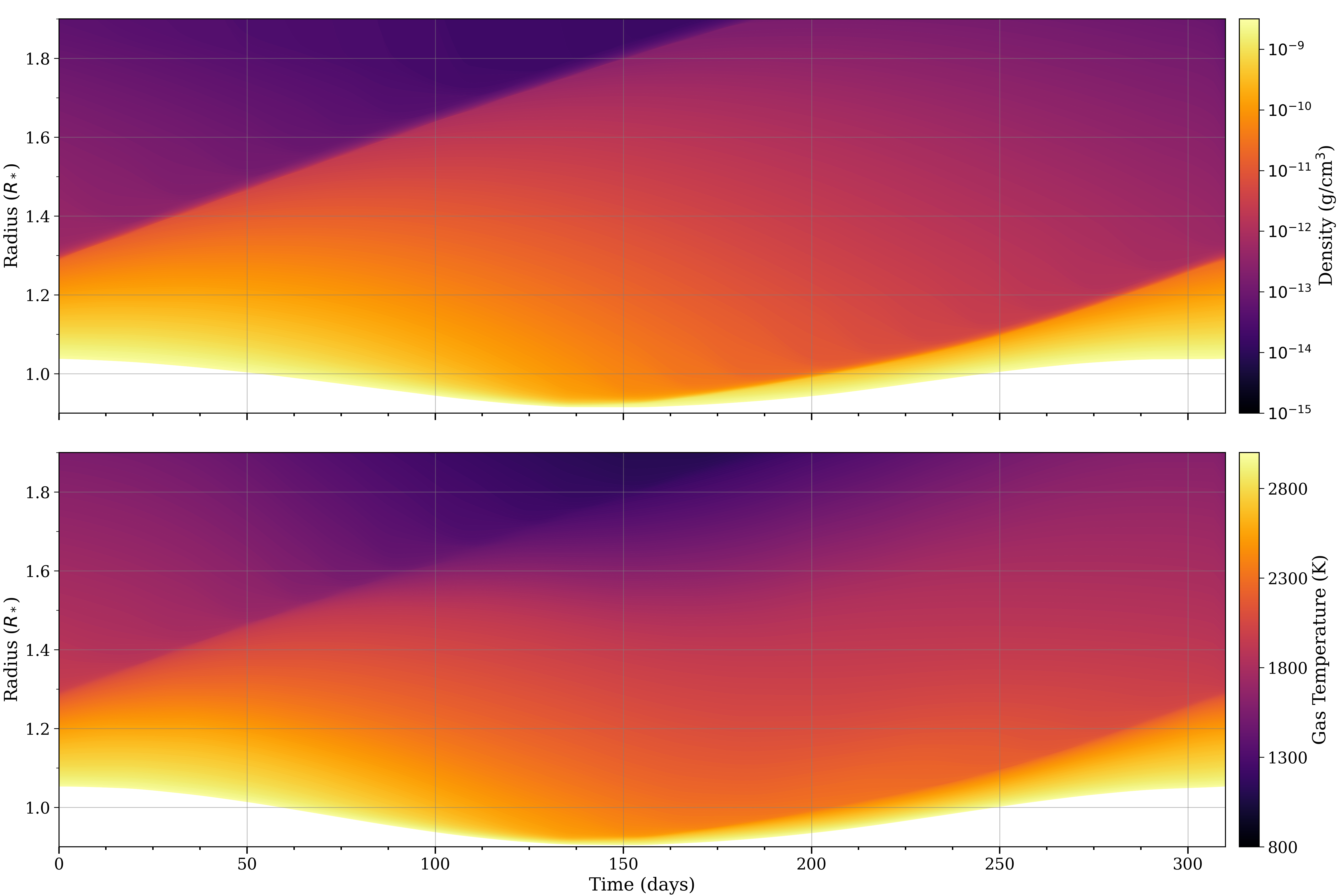}
        \caption{\textit{Top}: Density profile as a function of radial distance and time for the DARWIN atmospheric and wind model An315u3. The outward movement of two shock fronts is visible. \textit{Bottom}: Temperature profile as a function of radial distance and time for the same model. The bottom white area in both plots marks a region below the inner boundary of the model.}
        \label{fig:DARWIN_profile}
    \end{figure*}

    In this section, we describe our approach to modeling and synthesizing images of extended atmospheres around AGB stars. 
    We employ specific DARWIN models to calculate the appearance of the model star across wavelengths of interest throughout one complete pulsation cycle.

    \subsection{DARWIN models}
    \label{subsection:DARWIN models}

    \begin{table*}
        \centering
        \begin{minipage}{0.78\textwidth} 

        \caption{Input parameters and resulting wind properties of the DARWIN models}             
        \label{table:model_parameters}     
        \centering                         
        \begin{tabular}{l c c c c c c c c c c}   
            \hline\hline                 
            \multirow{2}{*}{Model} 
            & \multicolumn{1}{c}{$M_\star$} 
            & \multicolumn{1}{c}{$L_\star$}
            & \multicolumn{1}{c}{$T_{\rm eff}$}
            & \multicolumn{1}{c}{$R_\star$} 
            & \multicolumn{1}{c}{$P_\star$} 
            & \multicolumn{1}{c}{$\Delta u_P$} 
            & \multicolumn{1}{c}{$f_L$}
            & \multicolumn{1}{c}{$\Delta M_\textrm{bol}$}
            & \multicolumn{1}{c}{$\Delta V$}
            & \multicolumn{1}{c}{$\dot M$} \\
            & \multicolumn{1}{c}{($M_\odot$)} 
            & \multicolumn{1}{c}{($L_\odot$)} 
            & \multicolumn{1}{c}{(K)} 
            & \multicolumn{1}{c}{(AU)}
            & \multicolumn{1}{c}{(days)}  
            & \multicolumn{1}{c}{($km/s$)} 
            & \multicolumn{1}{c}{}
            & \multicolumn{1}{c}{(mag)}
            & \multicolumn{1}{c}{(mag)}
            & \multicolumn{1}{c}{($M_\odot/{\rm year}$)} \\
            \hline                        
            An315u3  & $1.0$ & $5000$ & $2800$ & 1.40 & $310$ & 3.0 & 2.0 & 0.54 & 1.9 & $3\times 10^{-7}$\\
            An315u4  & $1.0$ & $5000$ & $2800$ & 1.40 & $310$ & 4.0 & 2.0 & 0.73 & 3.8 & $4\times 10^{-7}$\\
            M2n315u6 & $1.5$ & $7000$ & $2600$ & 1.92 & $490$ & 6.0 & 1.5 & 0.95 & 6.2 & $2\times 10^{-6}$\\
            \hline                                   
        \end{tabular}
        \parbox{1\textwidth}{\footnotesize Note: Data from \citet{Hofner2022}. The radius $R_\star$ is calculated using the luminosity and effective temperature values. $\Delta u_P$ is the velocity amplitude at the inner boundary. $\Delta M_\textrm{bol}$ and $\Delta V$ are the bolometric and visual amplitude respectively. The parameters $f_L$ and $\Delta u_P$ determine the luminosity variations as shown in Eq. \ref{eq:Pulsation}.}
        \end{minipage}

    \end{table*}

    The DARWIN model\citep[Dynamic Atmosphere and Radiation-driven Wind models based on Implicit Numerics,][]{Hofner2016, Hofner2022} are one-dimensional radiation-hydrodynamical models designed to describe the time-dependent radial structures of evolved star atmospheres and winds.
    These spherically symmetric models encompass coupled systems of gas dynamics, radiative processes, and non-equilibrium dust formation. 
    The DARWIN models are particularly well-suited for studying radiative-transfer-dominated, optically thin atmospheres with complex molecular and dust chemistry \citep{Hofner2016}. 
    The inner boundary of these models is located just below the photosphere and above the driving zone of the pulsation. 
    
    In these models, the effects of stellar pulsation are simulated by prescribing periodic temporal variations in the radial position of the innermost mass shell with a velocity amplitude of $\Delta u_P$ and accompanying luminosity variations:
    
    \begin{equation}
    \label{eq:Pulsation}
        \begin{aligned}
        R_{\rm in}(t) &= R_0 + \frac{\Delta u_p P}{2 \pi} \sin \left( \frac{2 \pi}{P} t \right), \\
        L_{\text{in}}(t) &= L_0+ L_0 f_L \left( \frac{R_{\text{in}}^2(t) - R_0^2}{R_0^2} \right),
        \end{aligned}
    \end{equation}
    \citep[for details, see][]{Hofner2016}. 
    As is evident from these equations, the luminosity amplitude is defined by the changes in the radial position, together with an adjustable factor $f_L$. The resulting bolometric and visual amplitudes of the models are listed in Table \ref{table:model_parameters}. Starting from a hydrostatic, dust-free atmosphere, defined by the stellar parameters mass ($M_\star$), luminosity ($L_\star$), and effective temperature ($T_\textrm{eff}$), the DARWIN models are run for several hundred pulsation periods to avoid transients in dynamical behavior. 
    The radius $R_\star$ is the radius of the hydrostatic initial model, corresponding to the other stellar parameters listed in Table \ref{table:model_parameters}, computed using the equation $L_\star =4 \pi R_\star^2 \sigma T_\textrm{eff}^4$. It is a proxy for the time-averaged radius of the star, defined in the same sense.

    We use 21 snapshots of the dynamic model An315u3 to demonstrate the general results and models An315u4 and M2n315u6 to provide a broader perspective, with all three models introduced in \citet[]{Hofner2022}.
    These models include gradual Fe-enrichment of the wind-driving silicate dust, as described in detail in \citet{Hofner2022}.
    Table \ref{table:model_parameters} summarizes a selection of the stellar and atmospheric specifications. We selected model An315u3 as the main model due to its relatively low mass-loss rate and luminosity, which provide a reasonable approximation of the characteristics of stars R Leo and R Dor. Model An315u4, with its higher pulsation velocity amplitude, allows us to investigate the effects of increased radial pulsation amplitudes. Model M2n315u6 featuring a higher luminosity, larger mass, lower temperature, stronger wind, and longer period, allows for comparison under a different set of stellar conditions. The density and gas temperature profile of the model An315u3 are shown in Fig.~\ref{fig:DARWIN_profile} as a function of radius and time, and the corresponding profiles of An315u4 and M2n315u6 are illustrated in Fig.~\ref{fig:An315u4_profile} and Fig.~\ref{fig:M2n315u6_profile} respectively. The phases shown in the plots start at near-maximum bolometric luminosity, coinciding with the maximum radial extension of the stellar inner boundary, and cover one period. The variation of the bolometric luminosity is a good proxy for near-IR lightcurves (especially in the K-band), and also for variations in the visual regime, which are dominated by molecules like TiO, sensitive to atmospheric temperatures (see Höfner et al. 2022).

    \subsection{Radiative transfer}
    \label{subsection:Radiative transfer}

    \begin{figure*}
        \centering
        \includegraphics[width=0.7\linewidth]{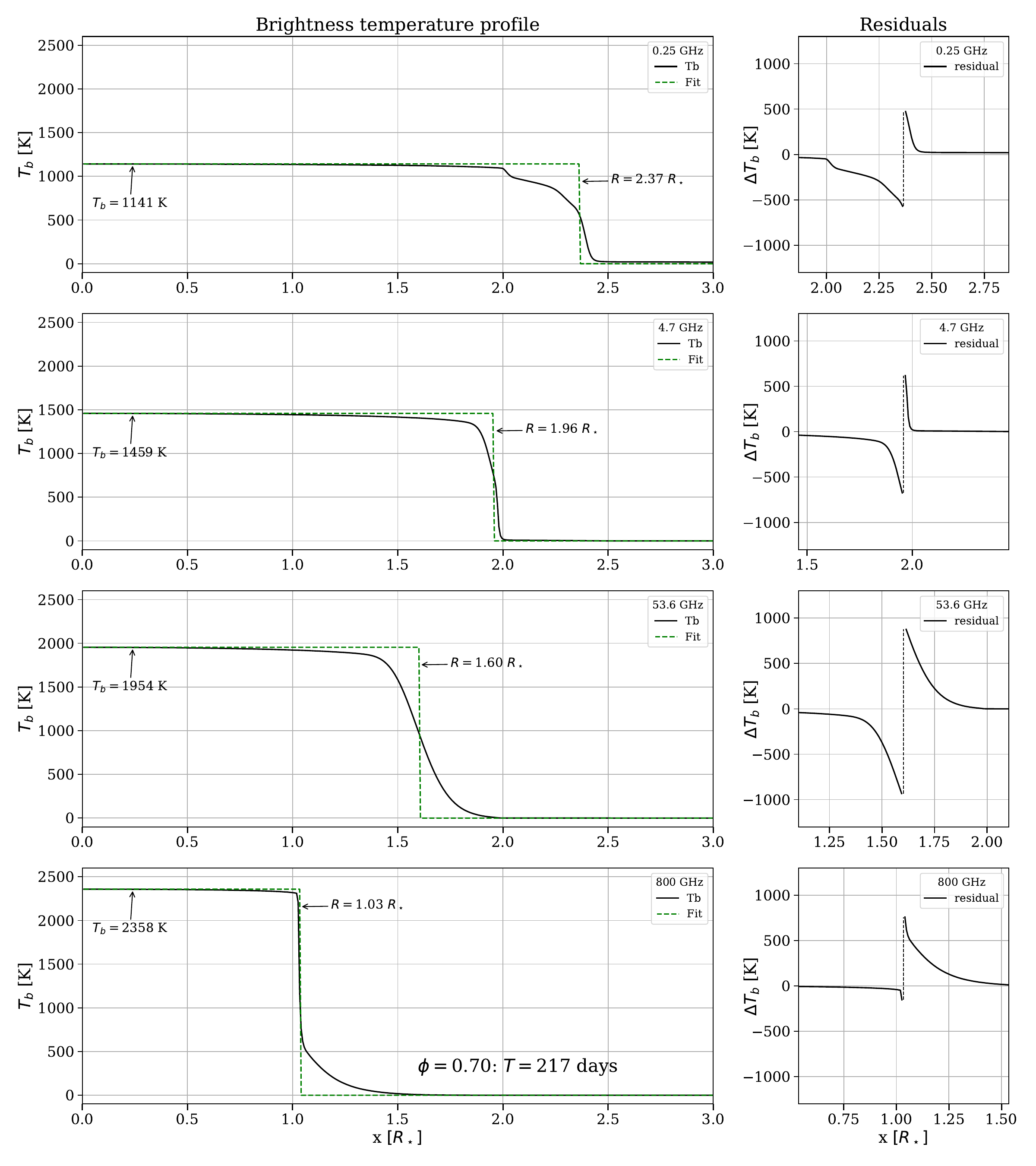}
        \caption{Brightness temperature profiles as a function of radial distance for four observation frequencies for model An315u3 at phase of 0.7 (217 days), with annotated fitted radii and disc brightness temperature. The fit residual values are illustrated on the right-hand plots, centered on the resulting radius from fitting. These four example frequencies were selected to demonstrate a range of different brightness temperature profiles observable simultaneously.}
        \label{fig:profile_residuals}
    \end{figure*}

    We create synthetic images at the required frequency to compare the models with observational data. Following \hyperlink{RM97}{RM97}, we considered electron-neutral free-free emission as the main opacity process within the temperature and density range of these models and relevant to radio and (sub)millimeter wavelengths.  
    To estimate dust opacity, we calculate the dust absorption cross section per mass using Mie theory and the optical constants for ${\rm Mg_2SiO_4}$ provided by \citet{Jaeger2003A&A}. We assume grain sizes of ${\rm 1~\mu m}$, but the particle size has virtually no effect on the calculated cross section per mass for particle sizes $\ll\lambda/2 \pi$. 
    Based on the gas density profile and the lowest gas-to-dust ratio of $\approx1200$ from the models, we find that at frequencies ${\rm \leq 1~THz}$, the radial optical depth due to dust is less than $10^{-3}$ for the model with the highest mass loss rate (M2n315u6, with mass loss of $2\times10^{-6}{\rm ~M_\odot/year}$), and even lower for other models and frequencies.
    
    \hyperlink{RM97}{RM97} assumed an equilibrium scenario with a timescale of less than 250 days within a dense radio atmosphere 
    ($>10^{-12}{\rm cm}^{-3}$)
    , where a mixture of molecular and atomic hydrogen was considered as the neutral particles for the resulting emission process.
    As shown in Fig.~\ref{fig:DARWIN_profile}, the DARWIN model in use drops quickly to lower densities in less than a stellar radius beyond the surface. Additionally, the timescales for local changes due to shocks (as seen in this model) and convective motions \citep[as seen in 3D models e.g.][]{Freytag2017} are shorter, which is also supported by observations \citep[e.g.][]{Khouri2016}. Consequently, following \hyperlink{V19}{V19}, we assume that the radio photosphere is not in aforementioned equilibrium and consider atomic hydrogen the dominant contributing neutral particle. 
    
    To estimate ionization, we assume local thermal equilibrium and apply Saha's equation \citep{Saha1920} at the relevant temperature ($T$) and electron density ($n_e$):
    \begin{equation}
    \label{eq:electron_density}
        \frac{n_{II}(j)}{n_I (j)}=\frac2{n_e} \frac{g_{II} (j)}{g_I(j)} \lambda^{-3} e^{-\Delta E(j)/kT},
    \end{equation}
    where $n_{II}(j)$ and $n_{I}(j)$ are respectively the singly-ionized and neutral number density of atomic species $j$, $g_{II}(j)$ and $g_{I}(j)$ are in order the degeneracy of ionized and neutral state of $j$, $\lambda$ is the de Broglie wavelength of an electron, $\Delta E(j)$ is the energy required to singly ionize the atom $j$, and $k$ is the Boltzmann constant.
    We assume solar abundances of elements \citep{Lodders2019} and following \hyperlink{RM97}{RM97}, adopt $K^+$, $Na^+$, $Ca^+$, and $Al^+$ as the sources of free electrons in the range of temperatures the models span.  Assuming these ions are the only contributors to the electron density in the environment (i.e., $n_e=\sum_j n_{II}(j)$), we calculate $n_e$ using a recursive optimization method.

    In order to obtain the opacity values, we utilize \hyperlink{RM97}{RM97} third-order polynomial fits for absorption coefficients at $10-{\rm GHz}$ frequency, and extend its frequency range using the frequency dependency evaluated by \citet{Dalgarno1966}:

    \begin{equation}
    \label{eq:kappa}
        \kappa_{\nu}=n_e k T n_H \nu^{-2}\sum_{i=0}^3{a_i \left(\frac{T}{10^3{\rm K}}\right)^i},
    \end{equation}
    where $a_0=3376$, $a_1=2149$ , $a_2=664.6$, and $a_3=78.53$.
    
    Integrating the absorption coefficient in the line of sight, we obtain frequency-dependent opacity values and solve the radiative transfer in the Rayleigh-Jeans limit to calculate brightness temperature $(T_b)$ profiles. 
    As seen in Figures \ref{fig:DARWIN_profile}, the shocks in the extended atmospheres cause abrupt changes in the temperature and density profiles. This pattern is also observed in the other two models, as depicted in Figures \ref{fig:An315u4_profile} and \ref{fig:M2n315u6_profile}. To mitigate computational errors in the radiative transfer, we use adaptive step sizes to keep the changes in brightness temperature under $0.1{\rm K}$ in each step. 
    
    We determine the star's central (disc) brightness temperature and photosphere radius by fitting a top-hat function (a disc model) to the $T_b$ profile of the stellar model. It's important to note that the values of the photosphere radius vary with frequency. Here, "photospheric radius" refers to the fitted radius of the brightness temperature at the given frequency. The residual of the $T_b$ profile after fitting is also of interest, as it allows a comparison with observational residual images (such as those shown in \hyperlink{V19}{V19}). Examples of brightness temperature profiles for four frequencies are illustrated in Fig.~\ref{fig:profile_residuals} together with the corresponding top-hat fits and residuals. We included further brightness temperature profiles for a wide range of frequencies and three phases for models An315u3, An315u4, and M2n315u6 in appendix Figs. \ref{fig:profile_residual_An315u3_phases}, \ref{fig:profile_residual_An315u4_phases}, and \ref{fig:profile_residual_M2n315u6_phases} respectively.

\section{Results}
\label{section:Results}

\begin{figure*}[htbp] 
  \centering
  \begin{minipage}[b]{0.49\textwidth}
        \centering
        \includegraphics[width=\textwidth]{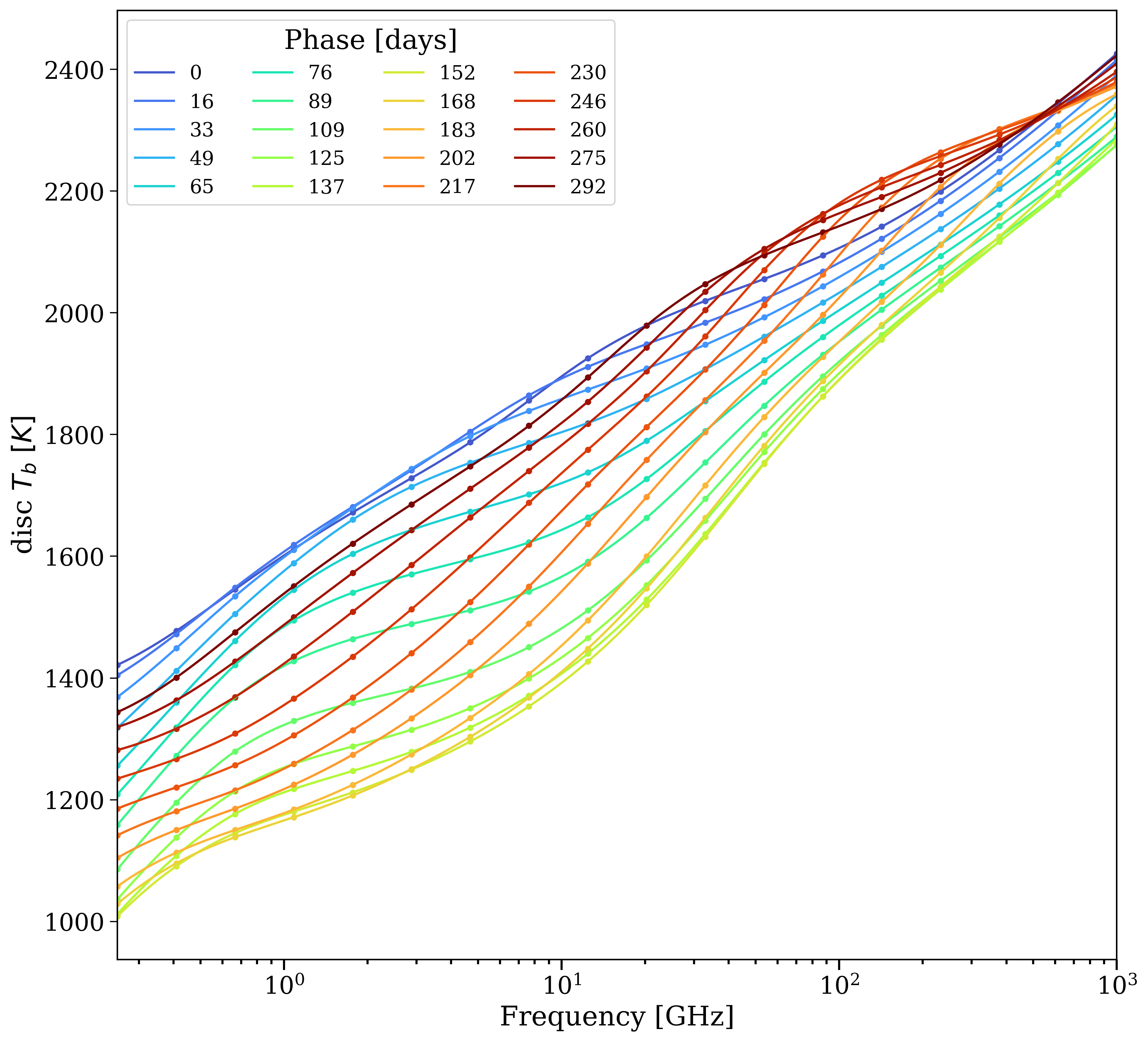}
        \label{fig:disc brightness temperature}
      \end{minipage}
      \hfill
      \begin{minipage}[b]{0.49\textwidth}
        \centering
        \includegraphics[width=\textwidth]{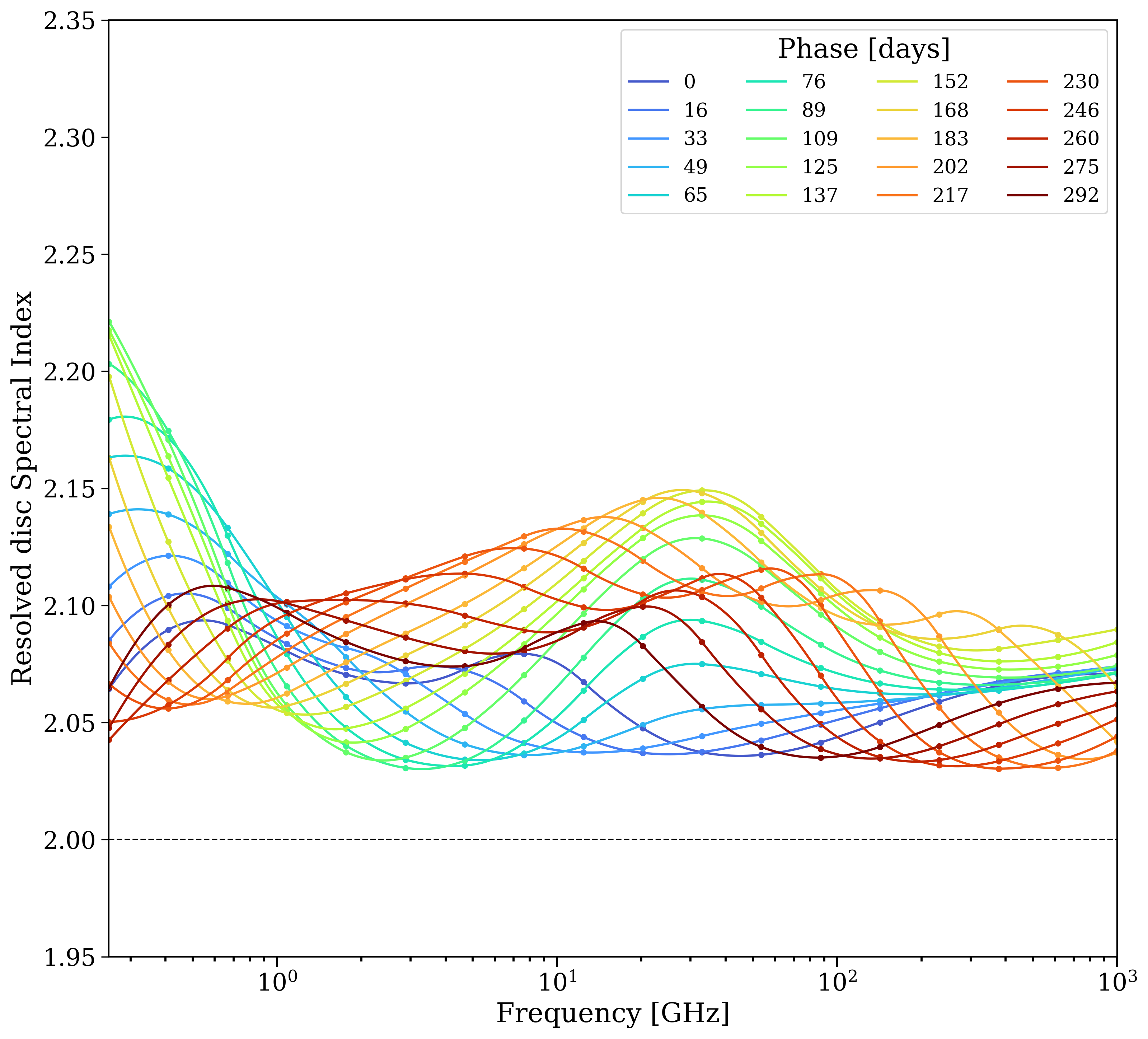}
        \label{fig:resolved spectral index}
      \end{minipage}

  \vspace{\floatsep} 
  
  \centering
  \begin{minipage}[b]{0.49\textwidth}
    \centering
    \includegraphics[width=\textwidth]{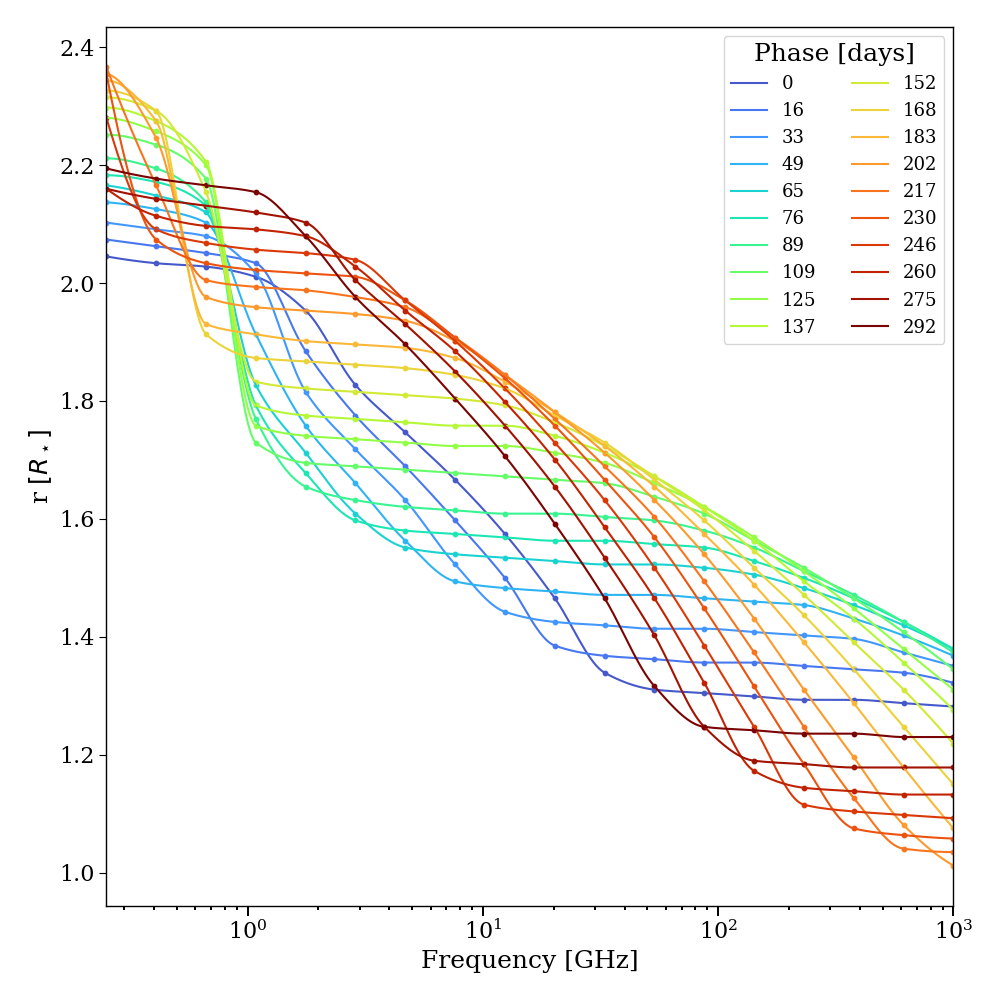}
    \label{fig:radius-freq}
  \end{minipage}
  \hfill
  \begin{minipage}[b]{0.49\textwidth}
    \centering
    \includegraphics[width=\textwidth]{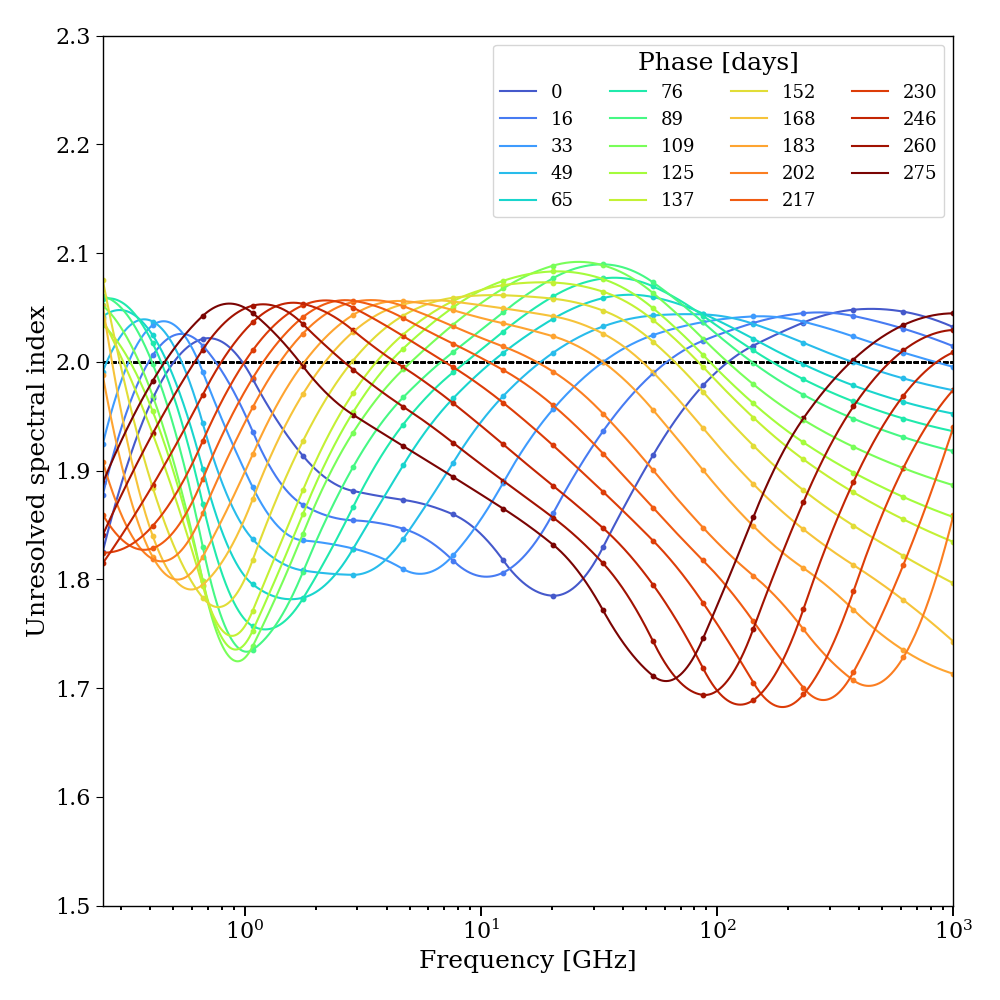}
    \label{fig:unresolved spectral index}
  \end{minipage}
  
  \caption{Observable quantities as a function of frequency based on synthesized observations for 21 snapshots of model An315u3 within a pulsation cycle. The frequency range encompasses bands from ALMA and VLA, as well as proposed ngVLA and SKA bands. \textit{Top left}: disc brightness temperature for resolved observations. \textit{Top right}: Resolved disc spectral index calculated from the disc brightness temperature. A dashed black line is used to indicate a spectral index of 2. \textit{Bottom left}: Photospheric radius. \textit{Bottom right}: Spectral index calculated from the flux density output for unresolved observations. Again, a dashed black line is used to indicate a spectral index of 2.}
  \label{fig:all_plots}
\end{figure*}

\begin{figure*}
    \centering
   \begin{subfigure}
        \centering
        \label{subfig:discTb-T}
        {\includegraphics[width=0.95\linewidth]{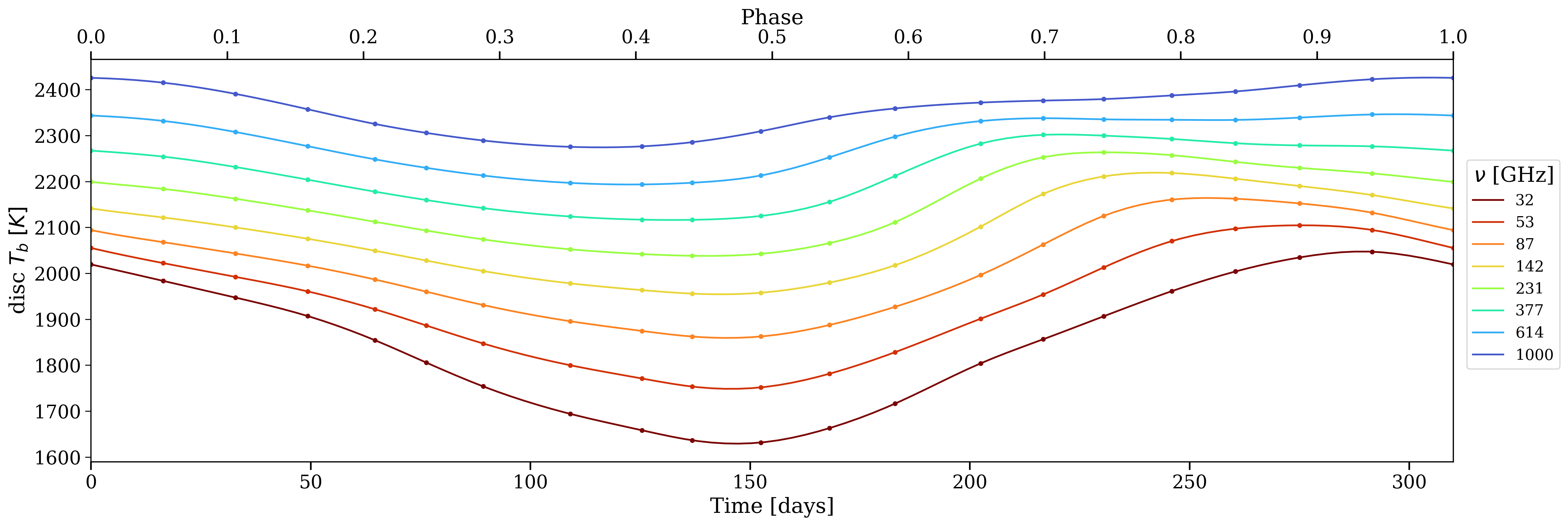}}
   \end{subfigure}
   
   \begin{subfigure}
        \centering
        \label{subfig:radius-T}
       {\includegraphics[width=0.95\textwidth]{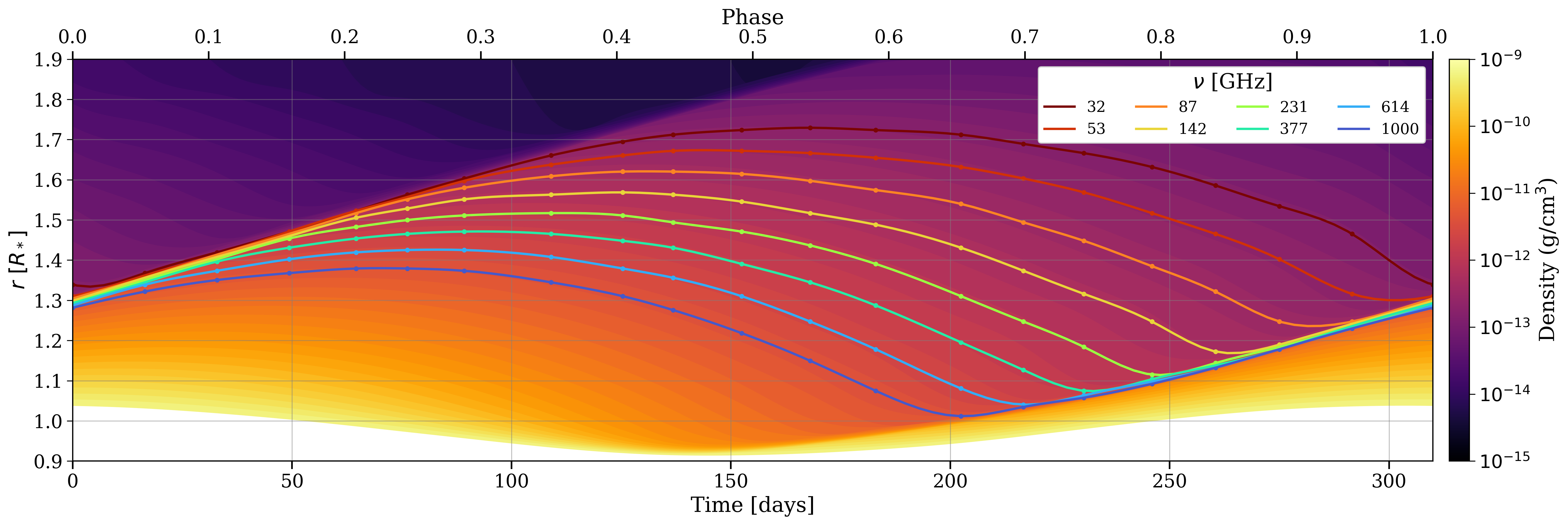}}
   \end{subfigure}
   
   \begin{subfigure}
        \centering
        \label{subfig:FluxDensity-T}
       {\includegraphics[width=0.95\textwidth]{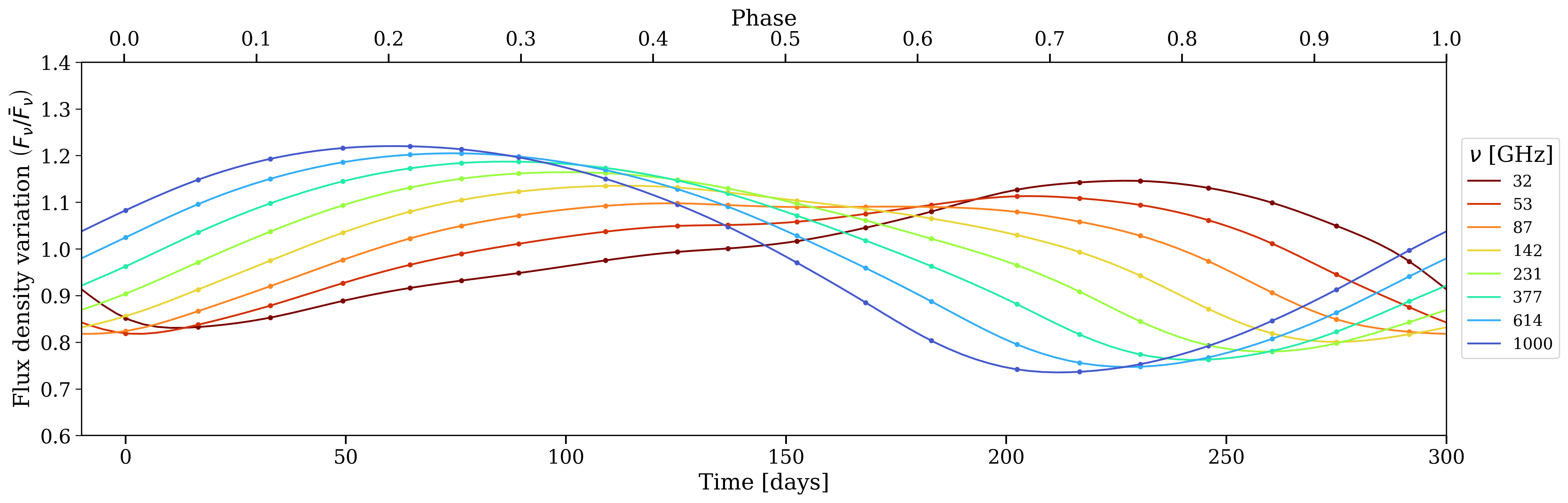}}
   \end{subfigure}

    \caption{Observable quantities for model An315u3, depicted as a function of time within one period across 8 frequencies within the range of ALMA bands, with the period spanning approximately 310 days, are shown with the corresponding luminosity phase indicated at the top of each plot. 
    \textit{Top}: disc brightness temperature variations. 
    \textit{Middle}: Variations in photosphere radii represented as plotted lines. The density profile of the stellar atmosphere, akin to Fig.~\ref{fig:DARWIN_profile}, serves as the background. Here, a visibly discrete color map is employed to enhance the visualization of the relationship between the photosphere radii and changes in the density profile.
    \textit{Bottom}: Flux density divided by the average flux density over one period.}
\label{fig:temporal_changes}
\end{figure*}

\subsection{Brightness temperature and resolved spectral index}
\label{subsection:Brightness temperature and resolved}
As shown in Fig.~\ref{fig:profile_residuals}, the brightness temperature profiles at different frequencies generally consist of a flat plateau followed by a transition slope at the edge of the stellar disc. Because of this wide plateau, the disc brightness temperature derived from the fitted top-hat function serves as a good approximation for the central beam brightness temperatures in observations of these models with resolved beams at least 2-3 times smaller than the stellar disc diameter (we call observations with less than 2-3 beams across the disc as marginally resolved).
In Fig. \ref{fig:all_plots}, we illustrate the frequency-related behavior of four observable parameters from the synthetic observation of different phases of the An315u3 model: top left plot depicts the stellar disc brightness temperature profile, while the bottom left plot shows the radius profile, both of which are the outcome of examining the $T_b$ profile as explained in Subsection \ref{subsection:Radiative transfer}. The resolved spectral index (top right) will be explained in this section, and the unresolved spectral index (bottom right) will be discussed in Subsection \ref{subsection:Flux density and unresolved}.

As depicted in Fig.~\ref{fig:all_plots} (top left), for the synthetic images of instances from model An315u3, the stellar disc brightness temperature $T_b^{\rm disc}$ (in the line of sight towards the stellar center) maintains a positive slope across frequencies, regardless of the phase of the model, a trend also observed in power law models (as seen in \hyperlink{RM97}{RM97} and \hyperlink{V19}{V19}). 
This consistent positive slope is attributed to the decreasing absorption coefficient and subsequent reduction in opacity with increasing frequency, as outlined in Eq. \ref{eq:kappa}, which results in migration of the characteristic region where the continuum is produced to lower radii, which is generally hotter for this model. Conversely, in models featuring regions where the gas temperature increases as a function of radius, higher frequency observations can result in a decrease in the $T_b^{\rm disc}$ profile with frequency. This phenomenon is present in model An315u4, as illustrated in Fig.~\ref{fig:all_plots An315u4} (top left). The negative $T_b-\nu$ slope for the highest frequencies in late phases results from the enhanced gas temperature in the region $\approx~ 1 {\rm ~ R_\star}$ at phases close to 250 days, as shown in Fig.~\ref{fig:temporal_changes An315u4}. If observed, such an inversion could provide a measurement of the efficiency of cooling after the shock passage.

The resolved spectral index $\alpha$ is calculated using the brightness temperatures at the center of the disc:
\begin{equation}
\label{eq: spectral index}
    \alpha =\frac{\partial \log{(F_\nu)}}{\partial \log{(\nu)}} = 
    \frac{\partial \log{(T_b\nu^2)}}{\partial \log{(\nu)}} = 
    \frac{\partial \log{(T_b)}}{\partial \log{(\nu)}} +2,
\end{equation}
where we used the flux density ($F_\nu$) relation with brightness temperature assuming Rayleigh-Jeans approximation ($ F_\nu \propto T_b\nu^2 $).
This spectral index can be obtained empirically from spatially resolved observations. In models featuring a general decrease in gas temperature with radius, the resolved spectral index values consistently remain above 2 due to the increasing brightness temperature with frequency, as depicted in Fig.~\ref{fig:all_plots} (top right). This trend aligns with expectations for an optically thick medium characterized by outward-decreasing temperature and free-free emission (a value of 2 corresponds to optically thick emission at constant temperature). In Fig.~\ref{fig:all_plots An315u4} (top right), it is evident that for model An315u4, the resolved spectral index drops below 2 in cases where the observed brightness temperature decreases with increasing frequency.

The near-synchronous variations in disc brightness temperature across the discussed frequencies are evident in Fig.~\ref{fig:temporal_changes} (top), showing that the temperature variations during the pulsation cycle can be tracked in all frequencies.
Upon qualitative comparison with the temporal changes in gas temperature depicted in Fig.~\ref{fig:DARWIN_profile} (top), it becomes apparent that the variations in disc brightness temperature closely follow the gas temperatures in the extended atmosphere of the model star, confirming the same interpretation by \hyperlink{RM97}{RM97} and \hyperlink{V19}{V19}. We discuss the relation between brightness temperature and gas temperatures in Section \ref{subsection:brightness and gas temperatures correlation}.

\subsection{Photosphere radius}
\label{subsection:Photosphere radius}

\begin{figure*}[hbt!]
    \centering
    \includegraphics[width=1\linewidth]{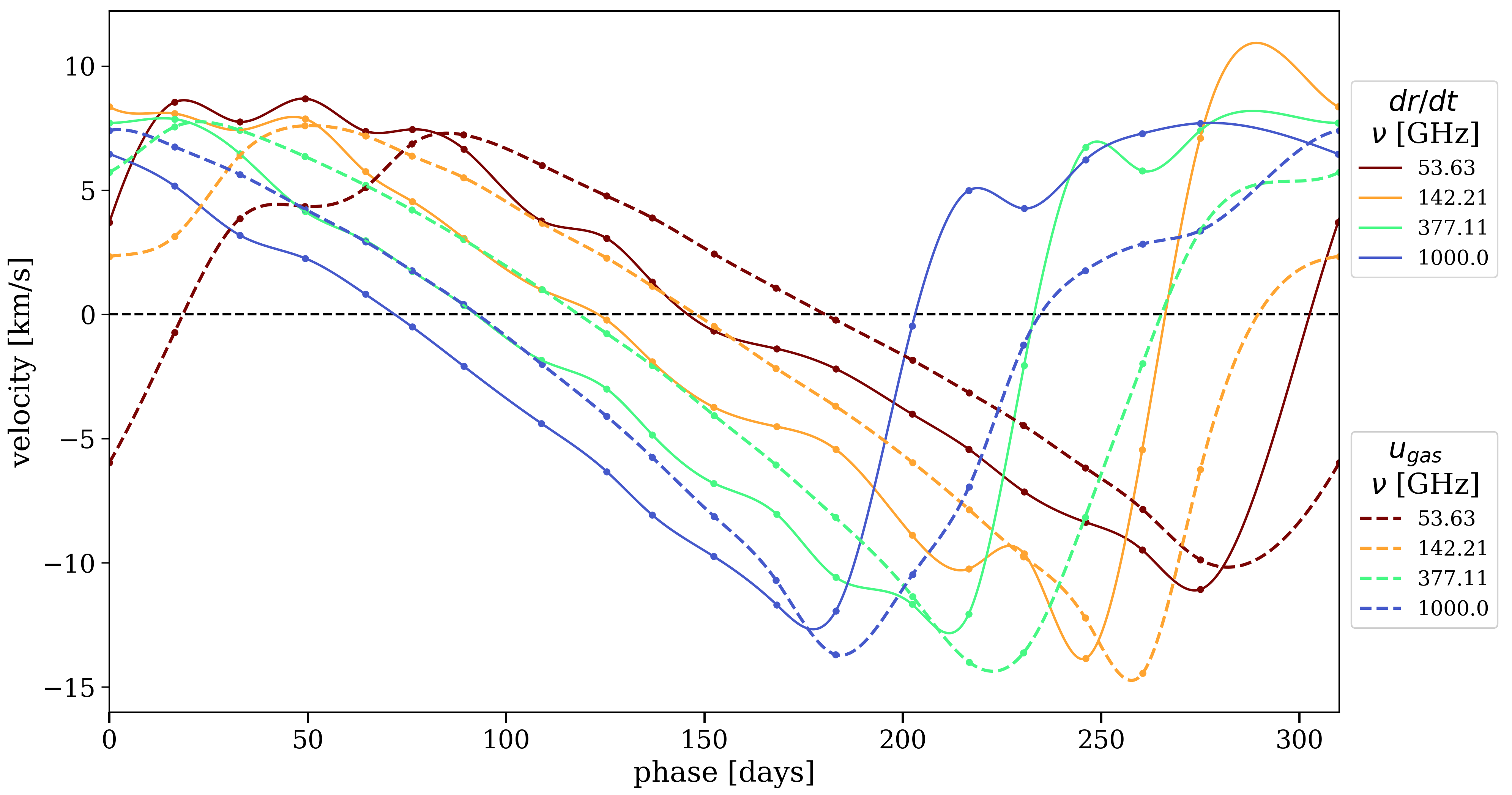}
    \caption{Surface velocity $dr/dt$ derived from the photospheric radius variations as a function of time plotted as lines for four frequencies within the range of ALMA bands for the star model An315u3. For comparison, the gas velocities $u_\text{gas}$ at the photosphere are plotted as dashed lines for the same phases.}
    \label{fig:surface_radial_velocity}
\end{figure*}

Fig.~\ref{fig:temporal_changes} (middle) exhibits the temporal variations in the photosphere radius of the model star within the ALMA frequency range, compared to the density profile of the models employed.
By comparing the density and temperature variations in Fig.~\ref{fig:DARWIN_profile}, it is evident that the density fluctuates more significantly, resulting in the measured radius primarily following the density profile.
The temporal evolution of radial profiles in this figure reveals two intervals distinguished based on the radial changes:
\begin{itemize}
    \item \textbf{Shock-driven expansion}: 
    During this phase, which lasts approximately half of the period, the shock-induced enhanced opacity layer results in consistent observed radii across a wide frequency range. This is depicted as a plateau in the radius-frequency plot (lower left panel of Fig.~\ref{fig:all_plots}), and the merging of radius-time profiles (middle panel of Fig.~\ref{fig:temporal_changes}). 
    This phenomenon can account for the relatively small differences in observed radii of the extended atmospheres of AGB stars in different frequencies when an expansion is observed (as seen in \hyperlink{V19}{V19}).
    Additionally, the steep transition slope of brightness temperature profiles during this phase leads to lower residuals (as shown in Fig.~\ref{fig:profile_residuals} and more extensively in Fig.~\ref{fig:profile_residual_An315u3_phases}).
    \item \textbf{Opacity layer contraction}: 
    As the shock expands to outer regions of the atmosphere and the density decreases, the radio photosphere decouples from the shock front, its expansion rate slows down, and eventually starts to contract. 
    The starting time of the transition between the shock-driven expansion and the contraction depends on the frequency: higher frequencies penetrate the shock front and reach the next shock earlier.
\end{itemize}

During the shock-driven expansion, the photospheric radius increases steadily with the outward-propagating shock front over time. Consequently, the expansion rate of the observed radio photosphere during this phase, indicated by the plateaus in the surface velocity profiles in Fig.~\ref{fig:surface_radial_velocity}, can be attributed to the shock’s expansion rate. On the other hand, while the changes in the measured radius do not precisely correspond to the gas velocities, the variation in the measured radii still reflects the range of gas velocities at the radio photosphere, as illustrated in Fig.~\ref{fig:surface_radial_velocity}.

Since the absorption coefficient is inversely proportional to the square of the frequency, observations at higher frequencies penetrate deeper into the extended atmosphere, resulting in smaller radii. 
We recover this expectation for the DARWIN models when calculating the radius as a function of frequency at a given pulsation phase. The trend is punctuated by intervals over which the radius is constant as a function of frequency, corresponding to 
the shock-driven expansion phase. 
As time progresses, the frequency range of the constant radius shifts to lower frequencies due to the diminishing density near the expanding shock.
At very low frequencies ($\nu\lesssim3$ GHz), the photosphere radii align with an outer shock front (which is a result of the previous pulsation), as depicted in Fig.~\ref{fig:radius-T extended}—which is the same as Fig.~\ref{fig:temporal_changes} but with frequencies extended to lower values—resulting in a notable increase in observed radii at these lower frequencies.

\begin{figure*}[hbt!]
    \centering
    \includegraphics[width=1\linewidth]{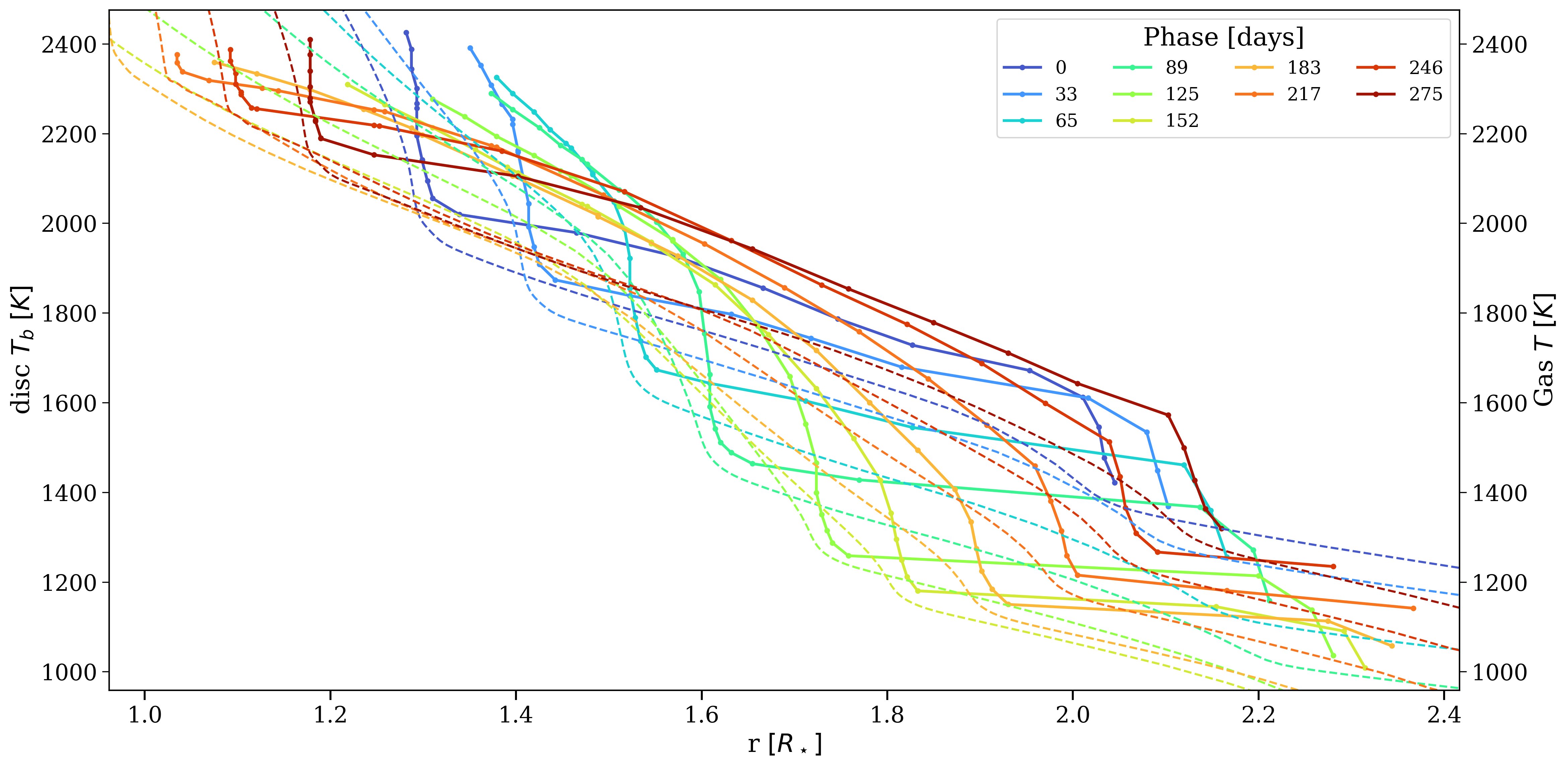}
    \caption{The brightness temperature values measured from synthetic imaging for 23 frequencies ranging from 250 MHz to 950 GHz, drawn for different phases in one period of the star model An315u3 as a function of radial distance. The dashed lines are gas temperature profiles for various phases as a function of radial distance.}
    \label{fig:CBT-radius}
\end{figure*}

\subsection{Flux density and unresolved spectral index}
\label{subsection:Flux density and unresolved}
We employ Rayleigh-Jeans approximation alongside the brightness temperature profile to compute the brightness intensity profile as a function of frequency. By integrating the obtained intensity profiles, we determine the flux density of the star as a function of frequency.
The resulting variation in flux density (flux density divided by the temporal-averaged flux density) within the frequency range of ALMA observations is depicted in Fig.~\ref{fig:temporal_changes} (bottom). This clearly illustrates that the flux density variation depends strongly on frequency.
As illustrated in Fig.~\ref{fig:temporal_changes} (top), the brightness temperature variations are delayed for lower frequencies. This delay corresponds to the heating progression through the atmosphere. On the other hand, the fitted-radius variations are considerably more delayed with lower frequencies, such that in the range of frequencies available for ALMA, the maximum observed radius at the lowest frequency corresponds to the minimum observed radius at the highest frequency. Given the flat-disc nature of the stellar brightness temperature profile from the DARWIN models, the unresolved flux density of the star is primarily influenced by the brightness temperature of the disc and the star's size at the specific frequency. Consequently, the different alignments between the radii and brightness temperature variations for different frequencies result in the inversion of the flux density variation within the frequency range available to ALMA.

We compute unresolved spectral index values using the flux density, which varies with frequency and pertains to unresolved observations.
In contrast to the resolved spectral index, which only depends on brightness temperature variations with frequency, the unresolved spectral indices depend on both the variations in brightness temperature and disc size across frequencies. Consequently, as illustrated in Fig.~\ref{fig:all_plots} (bottom right), the unresolved spectral indices fluctuate over values lower or higher than 2. For spectral indices measured in marginally resolved observations, we expect values between the resolved disc spectral index and the unresolved spectral index at the observed phase.

\subsection{Correlation Between Brightness and Gas Temperatures}
\label{subsection:brightness and gas temperatures correlation}

Disc brightness temperature values measured from fitting a disc model to resolved observations of the extended atmospheres are affected by the temperature and opacity profile of gas in the outer regions of the extended atmosphere. When an enhanced opacity layer like a shock front exists together with a low-opacity outer region, the temperature of the gas close to the enhanced opacity layer is dominant in the radiative transfer integration, which results in brightness temperatures close to the gas temperature values in the wake of the shock front.

In Fig.~\ref{fig:CBT-radius}, the disc brightness temperature values are illustrated as a function of photosphere radius for various frequencies and stellar phases within a period. The gas temperature is also shown as a function of radial distance to the stellar center. The brightness temperature profiles as a function of the photospheric radius exhibit similar values and slopes compared to the gas temperature profiles as a function of radial distance. Since the criteria of an enhanced opacity layer with a low-opacity outer region are satisfied for frequencies that reach the shock front but do not penetrate it, we conclude that the brightness temperature at that observation frequency closely resembles the local gas temperature. This frequency can be determined by the lowest frequency that results in the same photosphere radius in a multi-frequency observation. Utilizing this result can improve gas temperature estimations of shock fronts in the extended atmosphere, a crucial parameter for dust formation models. To measure the shock front brightness temperature at the lowest frequency, the observations should be carefully timed for the proper phase range, and the observation frequencies should be selected wide enough to detect the slope change in the radial profile of the disc brightness temperature.


\section{Discussion}
\label{section:Discussion}

Building upon the results presented in the previous section and motivated by our objective to evaluate evolved star models through observations in radio and mm/submm wavelengths, we now examine the capabilities of four current and upcoming radio telescope arrays: VLA, ALMA, SKA, and ngVLA. We aim to assess their potential contributions to advancing our understanding of the dynamics of extended atmospheres. The key specifications of the radio and mm/submm arrays used in this study are summarized in Table \ref{table:Arrays}.

\begin{figure*}[htbp] 
  \centering
  \begin{minipage}[b]{0.49\linewidth}
    \centering
    \includegraphics[width=\textwidth]{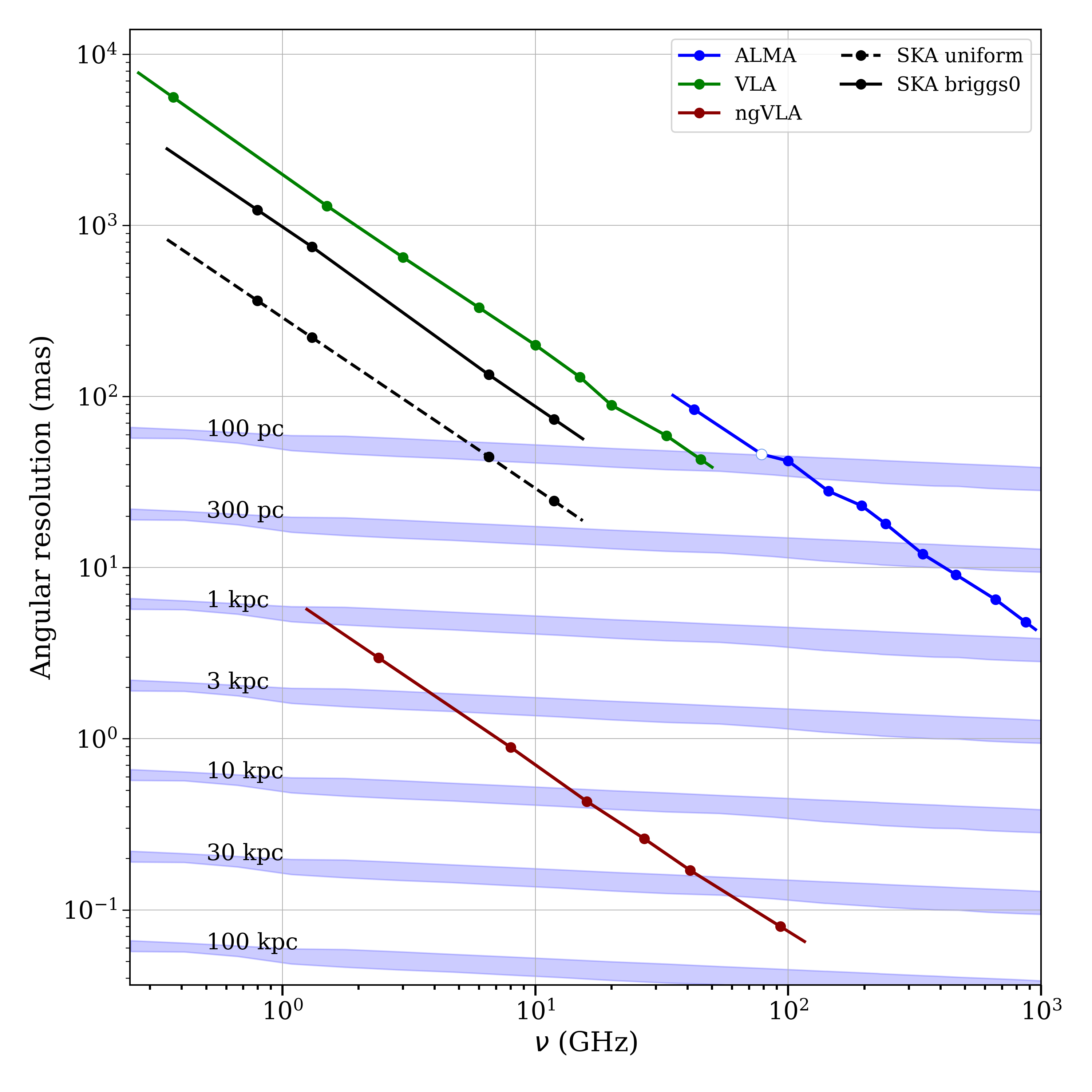}
    \label{fig:Angular_resolution_all}
  \end{minipage}
  \hfill
  \begin{minipage}[b]{0.49\linewidth}
    \centering
    \includegraphics[width=\textwidth]{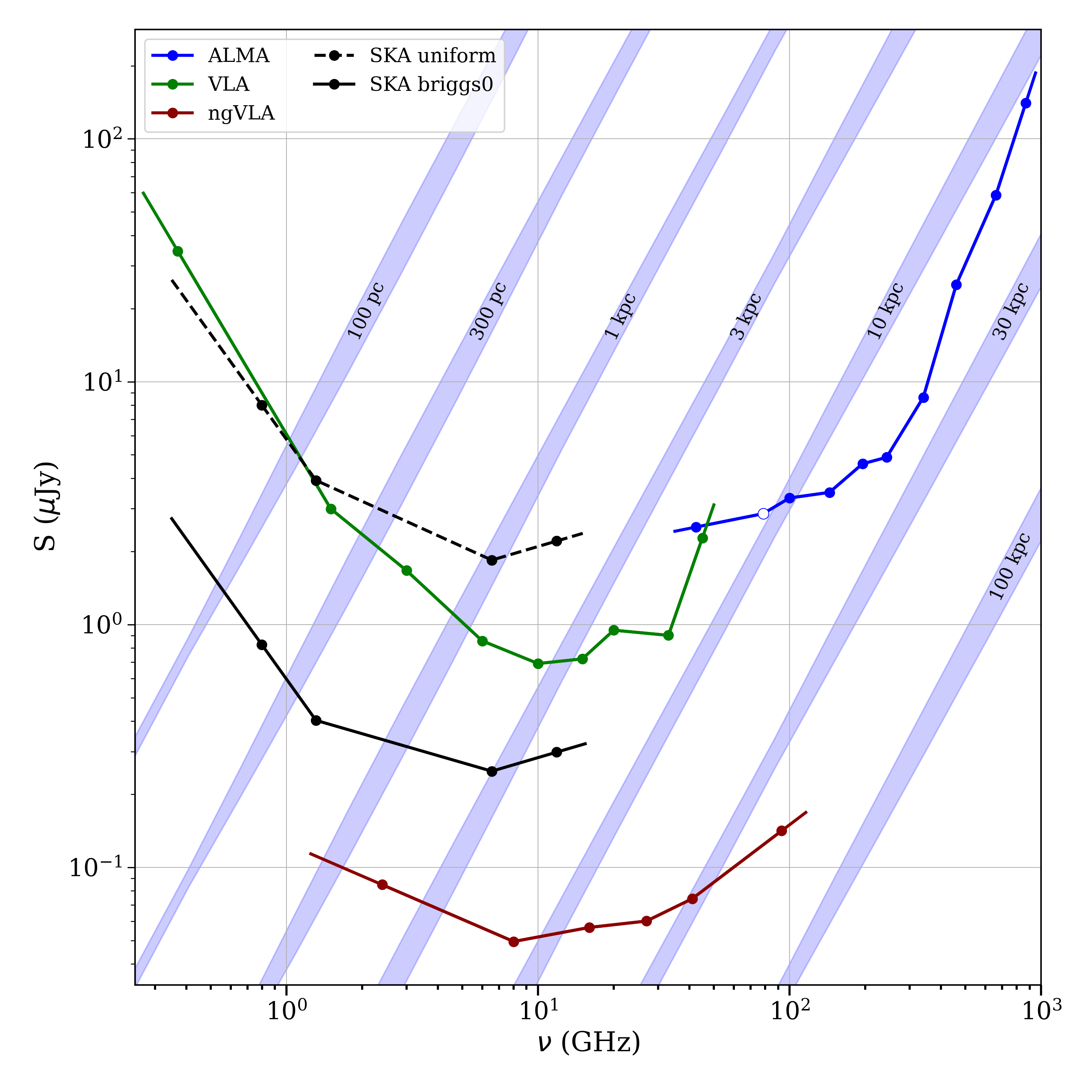}
    \label{fig:Flux_density_all}
  \end{minipage}
  \caption{Current and future observation capabilities (for an 8-hour integration) compared to the source distances as a function of observation frequency. \textit{Left}: Resolving power of current and future instruments as a function of frequency. The blue-filled areas show the range of frequency-specific resolved stellar angular diameters at different instances of the model An315u3 at various distances. \textit{Right}: Sensitivity power of current and future instruments as a function of frequency. The blue-filled areas show the frequency-specific stellar flux density range for unresolved observations of the instances of model An315u3 at various distances. The ngVLA sensitivity values are limited for resolutions lower than $10$~mas. 
  In both plots, the circles show the central frequencies of the respective band and the lines show the whole frequency range available to the telescopes. The hollow circle for ALMA shows the upcoming band 2.}
  \label{fig:Instruments_comparison}
\end{figure*}

 \begin{table*}
    \centering
    \begin{minipage}{0.8\textwidth} 
    \caption{Key properties of current and upcoming radio and mm/submm arrays}             
    \label{table:Arrays}     
    \centering                         
    \begin{tabular}{l c c c r}   
        \hline\hline                 
        Array   &Frequency range       &Antenna count &Antenna diameter &Longest baseline\\
        \hline                        
        VLA     &\phantom{0.}73 MHz - 50\phantom{.0} GHz &28   &25 m  &36.4 km\\
        ALMA    &\phantom{0.}35 GHz - 950\phantom{.} GHz &54   &12 m  &16   km\\
        SKA-mid &\phantom{.}350 MHz - 15.4\phantom{} GHz &133  &15 m  &150  km\\
        ngVLA Main   &\phantom{0}1.2 GHz - 116\phantom{.} GHz &214  &18 m  &1068 km\\
        ngVLA LBA   &\phantom{0}1.2 GHz - 116\phantom{.} GHz &30   &18 m  &8860 km\\
        \hline                                   
    \end{tabular}
    \end{minipage}
\end{table*}

The prominent radio frequency observatories for investigating the extended atmospheres of evolved stars include the Karl G. Jansky Very Large Array (VLA) in the northern hemisphere and the Atacama Large Millimeter/submillimeter Array (ALMA) in the southern hemisphere. The US-based VLA, with a four-decade history, operates in the frequency range of $73$~MHz to $50$~GHz, featuring 28 dishes of $25$~m diameter, and offers a configuration size range from $\textrm{1.0~km}$ to $\textrm{36.4~km}$. ALMA, located in Chile, is a cutting-edge radio telescope spanning frequencies from 35~GHz to 950~GHz. It boasts a flexible configuration with $54\times12$~m and $12\times7$~m dishes, enabling diverse setups ranging from 150m to 16km. We use sensitivity and resolution values relevant for VLA's full array with configuration A (36.4km baseline)\footnote{https://science.nrao.edu/facilities/vla/docs/manuals/oss/performance}. For ALMA, we consider the properties for the largest configuration (16~km) of $50 \times 12$~m dishes \footnote{https://almascience.eso.org/proposing/sensitivity-calculator}.

The Square Kilometer Array (SKA or SKA1) is a global initiative currently in development, comprising two telescopes: SKA-low, featuring 131,072 log-periodic dipole antennas across 512 aperture array stations in Australia, and SKA-mid, consisting of $133 \times 15$~m SKA dishes and $64 \times 13.5$~m Meerkat dishes in South Africa \citep[][and SKAO delivery plan]{Carilli2004, Braun2019}. SKA-low has a maximum baseline of 74 km and operates in the frequency range of 50 MHz to 350 MHz, while SKA-mid extends to a 150 km baseline with a frequency range of 350 MHz to 15.4 GHz.
The survey speed of SKA is supposed to reach two orders of magnitude higher than the current JVLA array
\citep{Braun2019}. In this study, we employed the SKA sensitivity calculator\footnote{https://sensitivity-calculator.skao.int} with the subarray configuration of AA4, an elevation of 45 degrees, and a water vapor column of $10$~mm.

The next generation Very Large Array (ngVLA) is a planned centimeter-wavelength observatory operating within the frequency range of 1.2-116~GHz, to be located in the USA, Mexico, and Canada. The facility comprises three subarrays: a main interferometric array housing $214 \times 18$~m diameter dishes with a baseline up to $\sim 1068 {~ \rm km} $, a short baseline array (SBA) featuring $19 \times 6$~m diameter dishes, and a long baseline array (LBA) equipped with $30 \times 18$~m diameter dishes extending to approximately 8860~km \citep{Selina2018}. 
Leveraging the continental baseline size of LBA, ngVLA can achieve resolutions of $\approx3$~mas in band 1 (2.4~GHz) frequency and lower than $0.1$~mas in band 6 (93~GHz) \citep{Murphy2018,Carilli2021NextGV}. For this study, we utilized observational properties of Revision D main+lba ngVLA subarrays, employing full-bandwidth natural weighting for the highest sensitivity limit\footnote{https://ngect.nrao.edu} and uniform weighting for the highest resolution\footnote{https://ngvla.nrao.edu/page/performance}.

\subsection{Unresolved observations}
\label{subsection:Unresolved observations}

We investigate the observational capabilities of current and future radio telescope arrays for unresolved observations of AGB stars by comparing the imaging sensitivity for long exposures for continuum observations with the flux density of the star across a broad spectrum of frequencies. 
The right panel of Fig. \ref{fig:Instruments_comparison} illustrates the sensitivity values of the four discussed arrays for an 8-hour integration time. For comparison, the flux density range of the mentioned instances of model An315u3 is plotted at various stellar distances. 
The two sets of values for SKA in both panels of Fig. \ref{fig:Instruments_comparison} demonstrate the range of SKA’s capabilities in terms of sensitivity and resolution under different weighting schemes.

Comparing the sensitivity of VLA across its range of bands with the expected flux density of the model shows that observing the nearest AGB stars with the VLA at low frequencies poses substantial challenges, bordering on impracticality due to the considerable investment of observation time required. Conversely, ALMA proves well-suited across all bands for efficiently monitoring the flux density of AGB stars in nearby galactic regions, allowing for a short observation duration for nearby sources.

The upcoming SKA-mid is poised to enhance imaging sensitivity within the frequency range covered by VLA by a factor of 2-4. The expansive field of view provided by SKA enables more frequent monitoring of nearby AGB stars, allowing for the generation of long-term time series data for a large number of sources. Moreover, the wide field of view and high sensitivity of SKA imply that AGB stars will often be detected in large-scale monitoring programs. Considering SKA-low, the flux density values from the model within the available frequency range are too low to be viable for observations with the array. The sizeable collecting area of the proposed ngVLA telescope substantially improves the sensitivity values, approximately doubling the scope of feasible unresolved observations compared to SKA, and extends the theoretical observation ranges in its high-frequency bands to encompass the edges of the Milky Way.

\subsection{Resolved Observations} 
\label{subsection:Resolved_observations}

To assess the capabilities of the current and future radio telescope arrays in resolving the extended atmospheres of evolved stars, we compare the nominal angular resolution of the arrays with the angular size of the model stars placed at different distances. As illustrated in Fig.~\ref{fig:Instruments_comparison} (left), the VLA demonstrates the capability to resolve our model star at distances within $\sim 100$~pc, particularly in the highest frequency bands. 
Recent observations with the VLA have aimed to resolve the extended atmospheres of nearby AGB stars, including W Hya, which was partially resolved by \hyperlink{RM97}{RM97}, as well as IRC+10216 \citep{Menten2012}, Mira \citep{Matthews2015}, R Leo, and $\chi$ Cyg \citep{Matthews2018}, all of which have been marginally resolved.

Despite its more compact configuration compared to VLA, ALMA extends to higher frequencies, consequently achieving superior angular resolution. With a resolution as fine as 5~milliarcseconds, ALMA can effectively resolve the extended atmospheres of AGB stars within $\approx 1~\textrm{kpc}$ (assuming the target stars are similar to our model), significantly expanding the number of observable sources by more than two orders of magnitude. Examples of recent observations include \citet{Matthews2015} resolving Mira, \citet{Vlemmings2018} resolving R Dor, and \hyperlink{V19}{V19} resolving the extended atmospheres of W Hya, Mira A, R Dor, and R Leo, including the measurement of an outward motion of the millimeter wavelength photosphere. Additionally, \citet[]{Vlemmings2024} resolved the stellar disc and small-scale surface structures on R Dor, along with the temporal variations of these surface features.

The upcoming SKA-mid bands extend the frequency range of resolving the nearest sources, complementing VLA and ALMA capabilities to achieve more than two orders of magnitude in frequency coverage, while SKA2, the future extension plan for SKA, is anticipated to attain resolutions comparable to ALMA \citep[][]{Braun2019}. The high sensitivity and wide field of view of SKA open up possibilities for incorporating resolved observations of extended atmospheres as an auxiliary goal in SKA's large programs. Conversely, the SKA-low bands, characterized by beam sizes exceeding one arcsec, are unsuitable for resolving the atmospheres of even the closest evolved stars.
Meanwhile, the envisioned capabilities of ngVLA promise galaxy-wide resolved observations of evolved stars and precise imaging of nearby stars, featuring up to approximately 100 beams across the stellar diameter.

As seen in the radius-frequency plot depicted in Fig.~\ref{fig:all_plots} (e.g. following the profiles for $\phi=230 - 292$ days) and also illustrated in Fig.~\ref{fig:radius-T extended}, the measured radii in simultaneous observations of the model star at frequencies available to ALMA and SKA/ngVLA reveal two distinct shock fronts at different radii, suggesting a promising opportunity to study the extended atmosphere layer by layer. 

Furthermore, monitoring radius changes across a broad frequency range enables the differentiation of outward shock fronts from contracting density profiles. 
This capability holds significant potential for directly mapping the structure of the extended atmosphere and rigorously testing the accuracy of the dynamical atmospheric models.
As mentioned in the previous section, ALMA is capable of resolving nearby AGB stars and is suitable for monitoring the radial movements of the optical surface. Resolved observations of nearby AGB stars can be utilized to constrain the stellar hydrodynamic models to measure the shock front velocities. As shown in Section \ref{subsection:Photosphere radius}, for a substantial time during an observed expansion, the shock front carries the optical surface at the observation frequency outward, and as a result, the radius variations during this phase represent the shock front velocity. 
Consequently, confirming the shock-driven expansion phase involves measuring identical radii across multiple frequency bands and observing a uniform increase in radii over time across these frequencies.

For instance, to track the shock front propagation in an AGB star akin to model A3n315u3 at a distance of 100 pc using the longest baseline of ALMA, we recommend scheduling closely timed observations in bands 7, 8, and 9 over approximately one month with observations spaced about one week apart. To monitor the radius variation profile, these observations should target late-phase ranges (day 250 to the end of the period) while the stellar bolometric phase can be inferred from previous IR observations. 
Including SKA and ngVLA in monitoring the radius of nearby AGB stars can extend the tracing of the shock front across all phases of the variation period. This extension enables the study of the extended atmospheres of AGB stars and the propagation of shocks from the region near the infrared photosphere up to approximately 2 stellar radii.

\subsection{Assumption of spherical symmetry}
The models used in this study are one-dimensional stellar atmospheric models, a simplification that may not fully capture the complex, asymmetric nature of AGB star’s extended atmospheres, as shown by the resolved observations \citep[e.g.][]{Matthews2018, Vlemmings2019, Vlemmings2024}.  These 1D models are valuable for comparing the average variations in the observational parameters of the star, providing insights into its overall behavior. Additionally, these models are instrumental in simulating slices radial structures of gas within the extensive convective cells. 

In the snapshots of the one-dimensional models presented, the shocks only show radial movement in the displayed radial distance range. 
Contrary to 1D models, lateral shock movements are present in 3D models \citep[e.g.][]{Freytag2023}, which can introduce further complications that are not covered in this study. Additionally, the shock fronts are not well-resolved in the current DARWIN models (and 3D models such as CO5BOLD), probably underestimating the peak temperatures in the shocks.


\section{Conclusion}
\label{section:Conclusion}

We generated synthetic images of Asymptotic Giant Branch (AGB) stars' extended atmospheres using both simple power-law models and DARWIN one-dimensional atmospheric and wind models in radio and mm wavelengths. By analyzing the brightness temperature profiles, we derived key observable features, including photosphere radius (radius of the fitted disc), disc brightness temperature (central brightness temperature of the brightness temperature profile), and resolved spectral index (from disc brightness temperature profile as a function of frequency), and unresolved spectral index (from unresolved flux density). Our examination uncovered variations in observable features that depend on frequency and phase. Notably, the radius-frequency profile of the model star reveals frequency-specific ranges of phase with consistent measured radii. This pattern indicates the presence of layers with heightened continuum opacity attributed to high densities behind shock waves. We demonstrate that during the shock-driven expansion phase, the observed increase in stellar radius follows the expanding density shock front. Conversely, in the contraction phase, we establish that the measured radius tracks the gas density profile, decreasing as the gas is diluted between successive shock fronts.
Additionally, our findings indicate that models with strong shocks present an inversion in the brightness temperature as a function of frequency that might be observable in the highest frequencies available to ALMA. This will allow the observations to constrain the cooling efficiency behind the shock.

We assessed the observational capabilities of current and upcoming radio/mm arrays, including VLA, ALMA, SKA, and ngVLA, for detecting and resolving AGB stars. The high imaging sensitivity and wide field of view of SKA expand the observability range of AGB stars and extend the frequency range for studying the extended atmospheres of nearby sources. Furthermore, the higher angular resolution of SKA-mid proves effective in resolving the nearest sources in centimeter wavelengths, enabling resolved observation of separate layers of shock fronts in simultaneous observations with ALMA. We also discussed how ngVLA’s proposed high sensitivity and angular resolution can bring the study of extended atmospheres of AGB stars to a Galaxy-wide range and can resolve the nearest sources in an unprecedented degree of detail.

While our study offers valuable insights into the current and future observation of AGB stars in radio and mm/submm wavelengths, it is essential to acknowledge certain limitations that may influence the generalization of the findings. The models used in this study are not suited for modeling the asymmetries and lateral in the synthetic observations. 
Future works should compare our findings with analyses using 3D AGB star atmosphere models. 

The findings presented in this study pave the way for a more comprehensive understanding of the atmospheres of AGB stars in radio and mm/submm wavelengths. As advancements in observational capabilities and modeling techniques continue, we anticipate increasingly accurate portrayals of the intricate dynamics governing AGB stars' atmospheres.

\section*{Acknowledgments}
We thank Kjell Eriksson for his help regarding the relationship between pulsation and brightness phases in the models. 
B.B., W.V., and T.K. acknowledge VR support under grants No. 2019-03777 (TK) and 2020-04044 (BB and WV), as well as support from the Olle Engkvist foundation under project 229-0368. 
S.H. acknowledges funding from the European Research Council (ERC) under the European Union’s Horizon 2020 research and innovation program (Grant agreement No. 883867, project EXWINGS) and the Swedish Research Council (Vetenskapsradet, grant number 2019-04059).

\vspace{5mm}
\clearpage
\appendix

\section{Additional DARWIN models and extended frequencies}
\label{appendix:Additional DARWIN models}

\begin{figure}[ht]
    \centering
    \rotatebox{0}{%
        \includegraphics[width=0.9\textwidth]{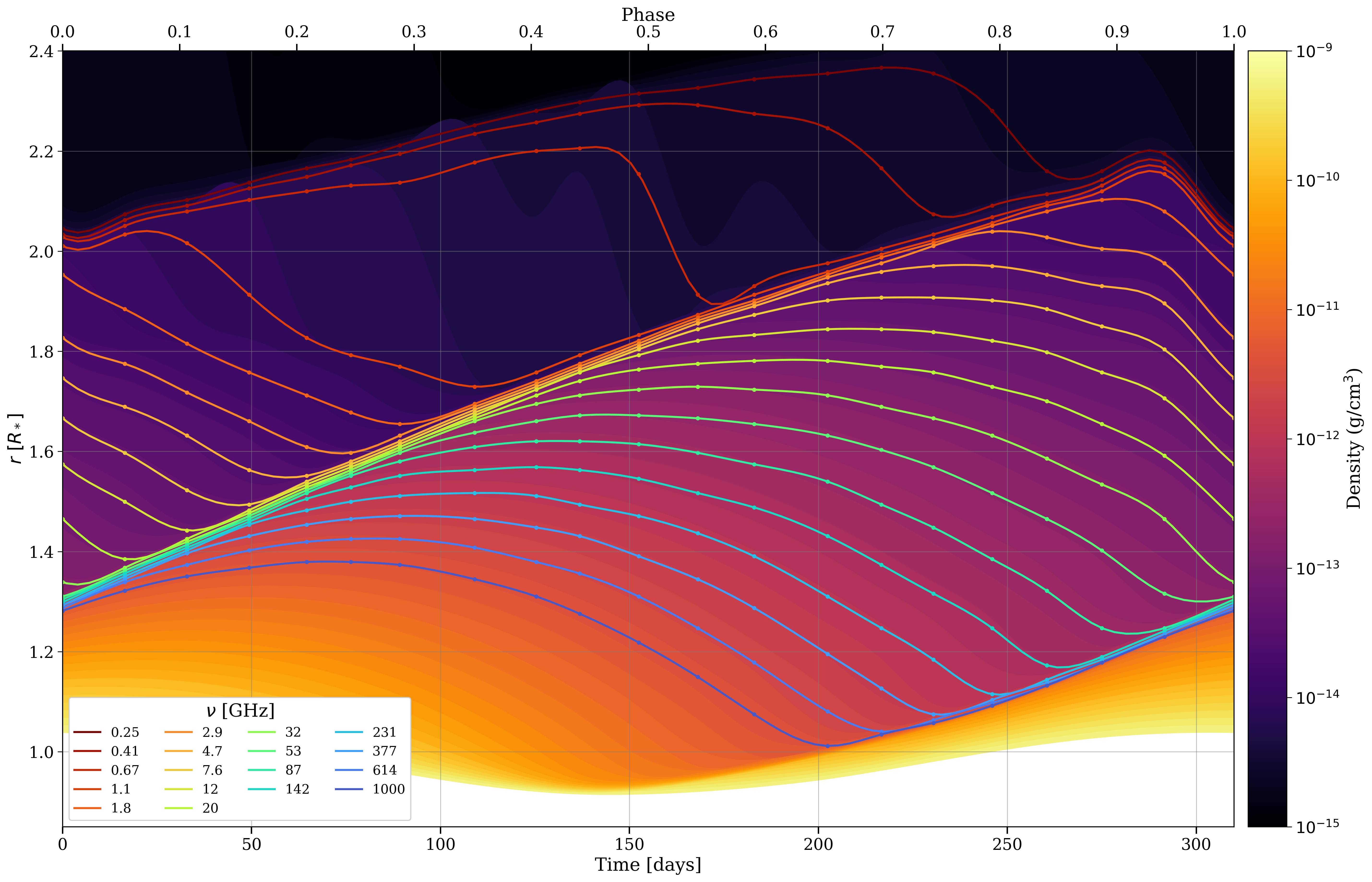}
    }
    \caption{Same as Fig.~\ref{fig:temporal_changes} (mid) with extended frequency range down to 250 MHz.}
    \label{fig:radius-T extended}
\end{figure}

\begin{figure}[hb]
    \centering
    \includegraphics[width=0.9\linewidth]{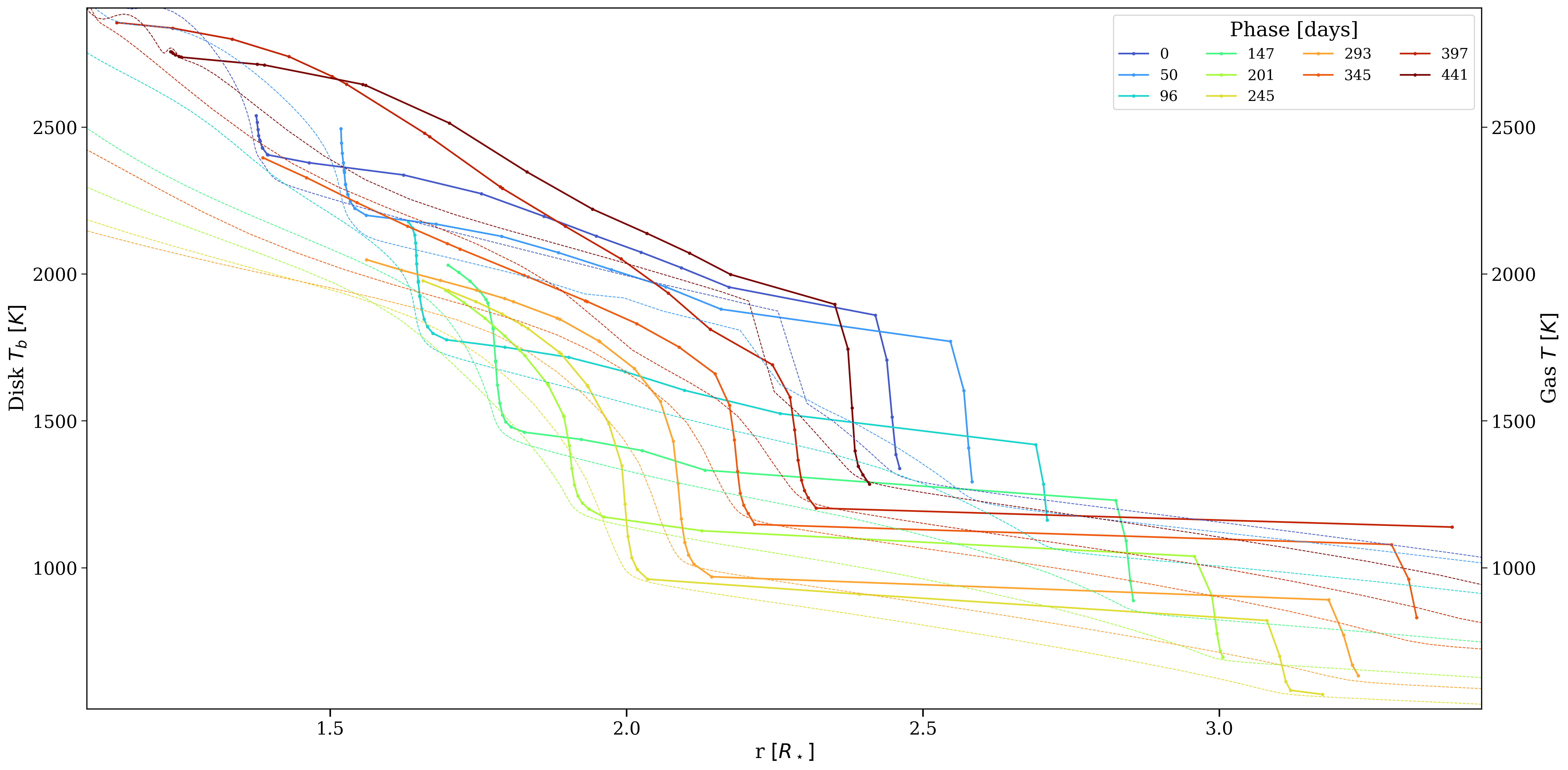}
    \caption{Same as Fig.~\ref{fig:CBT-radius} for model M2n315u6.}
    \label{fig:CBT-radius M2n315u6}
\end{figure}

We present the temporal variation in the photosphere radius of the model An315u3 with an extended frequency range down to ${\rm 250~MHz}$ in Fig. \ref{fig:radius-T extended}, illustrating the observation of two consecutive shock fronts in simultaneous observations with ALMA and SKA/ngVLA. We show disc brightness temperature values as a function of the photospheric radii for the same range of frequencies as used in Fig. \ref{fig:all_plots} for the model M2n315u6 in Fig. \ref{fig:CBT-radius M2n315u6}.

\begin{figure*}
    \centering
    \includegraphics[width=0.85\linewidth]{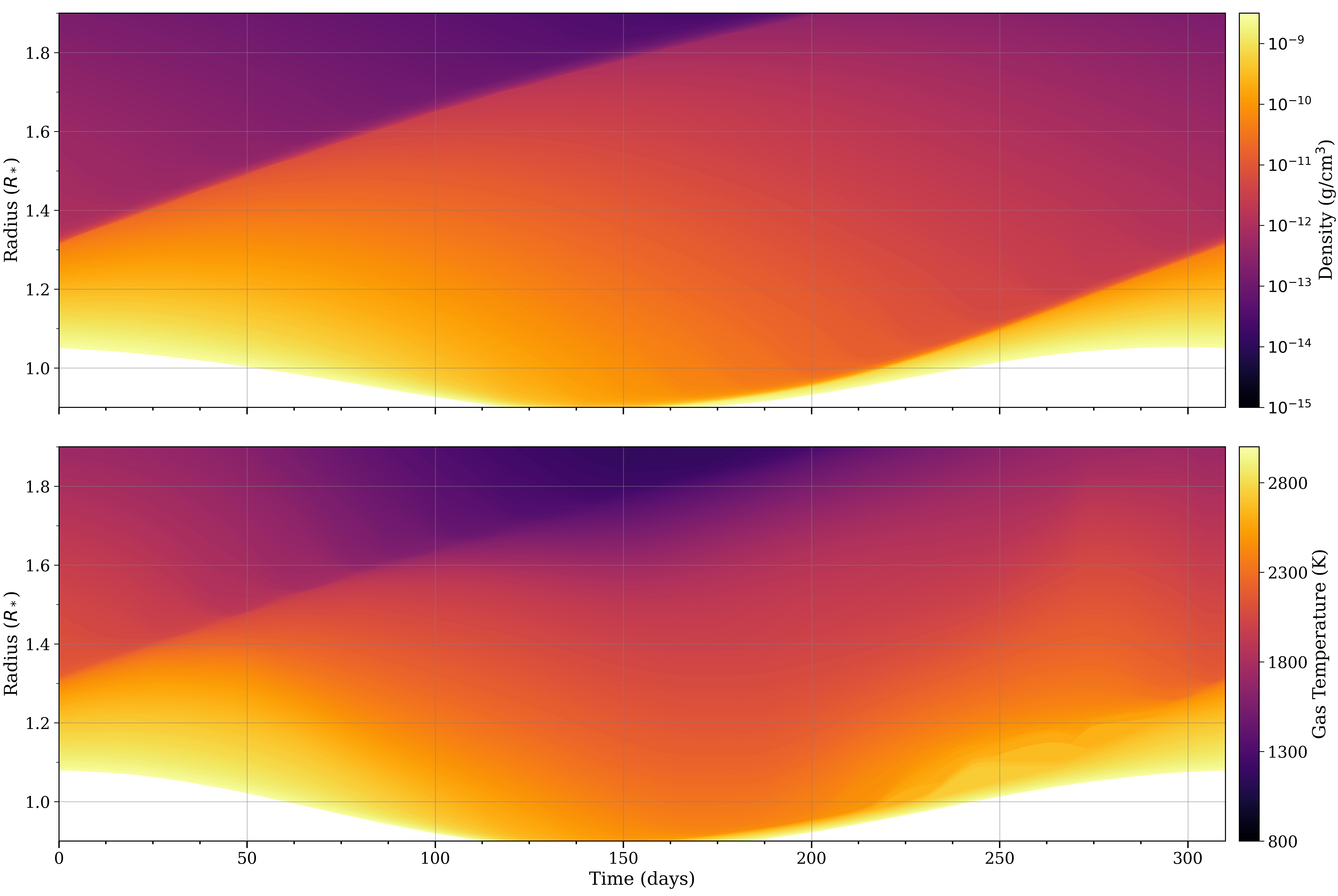}
    \caption{Same as Fig.~\ref{fig:DARWIN_profile} for model An315u4.}
    \label{fig:An315u4_profile}
\end{figure*}

\begin{figure*}[htbp] 
  \centering
  \begin{minipage}[b]{0.49\textwidth}
        \centering
        \includegraphics[width=\textwidth]{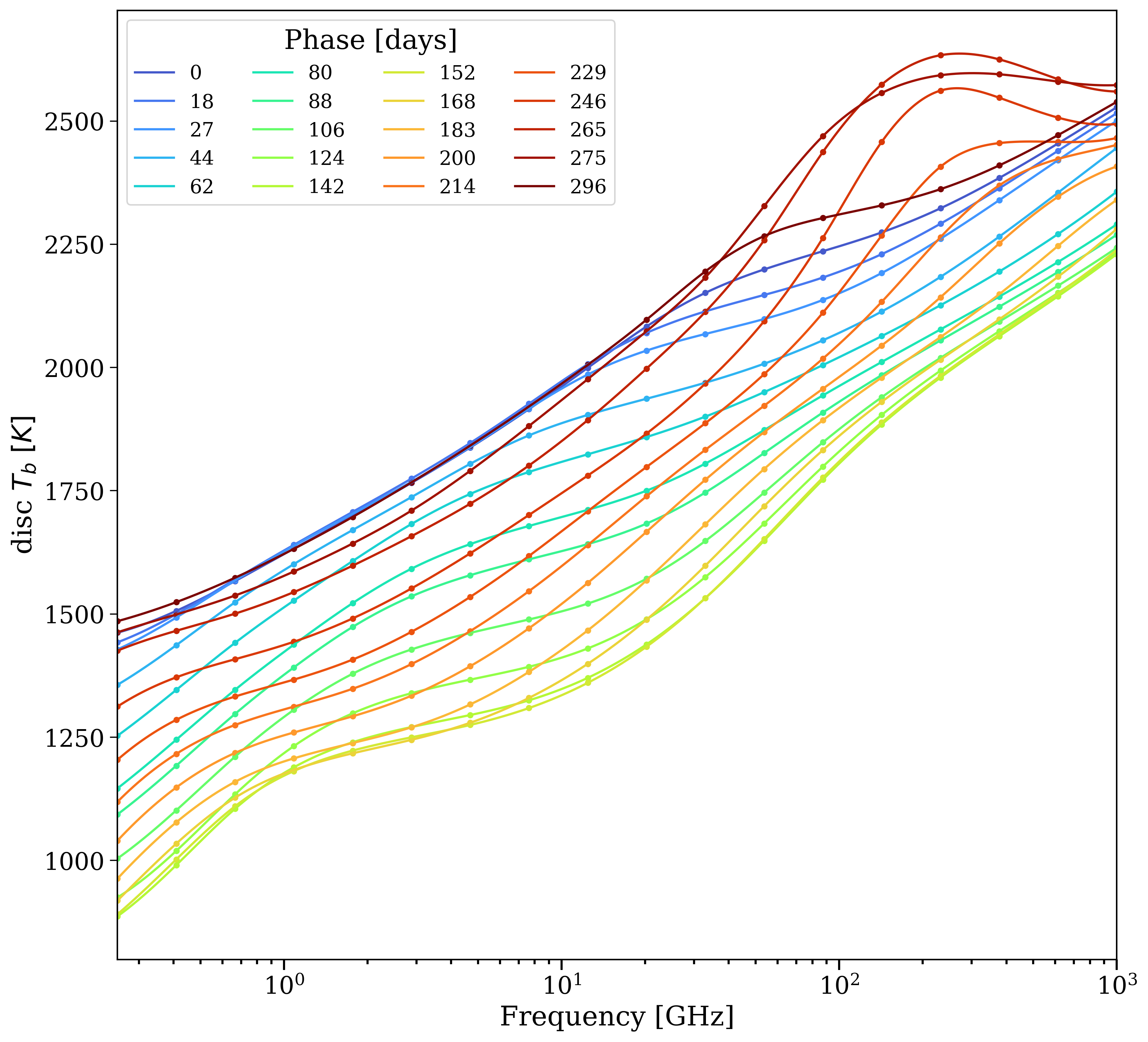}
        \label{fig:disc brightness temperature An315u4}
      \end{minipage}
      \hfill
      \begin{minipage}[b]{0.49\textwidth}
        \centering
        \includegraphics[width=\textwidth]{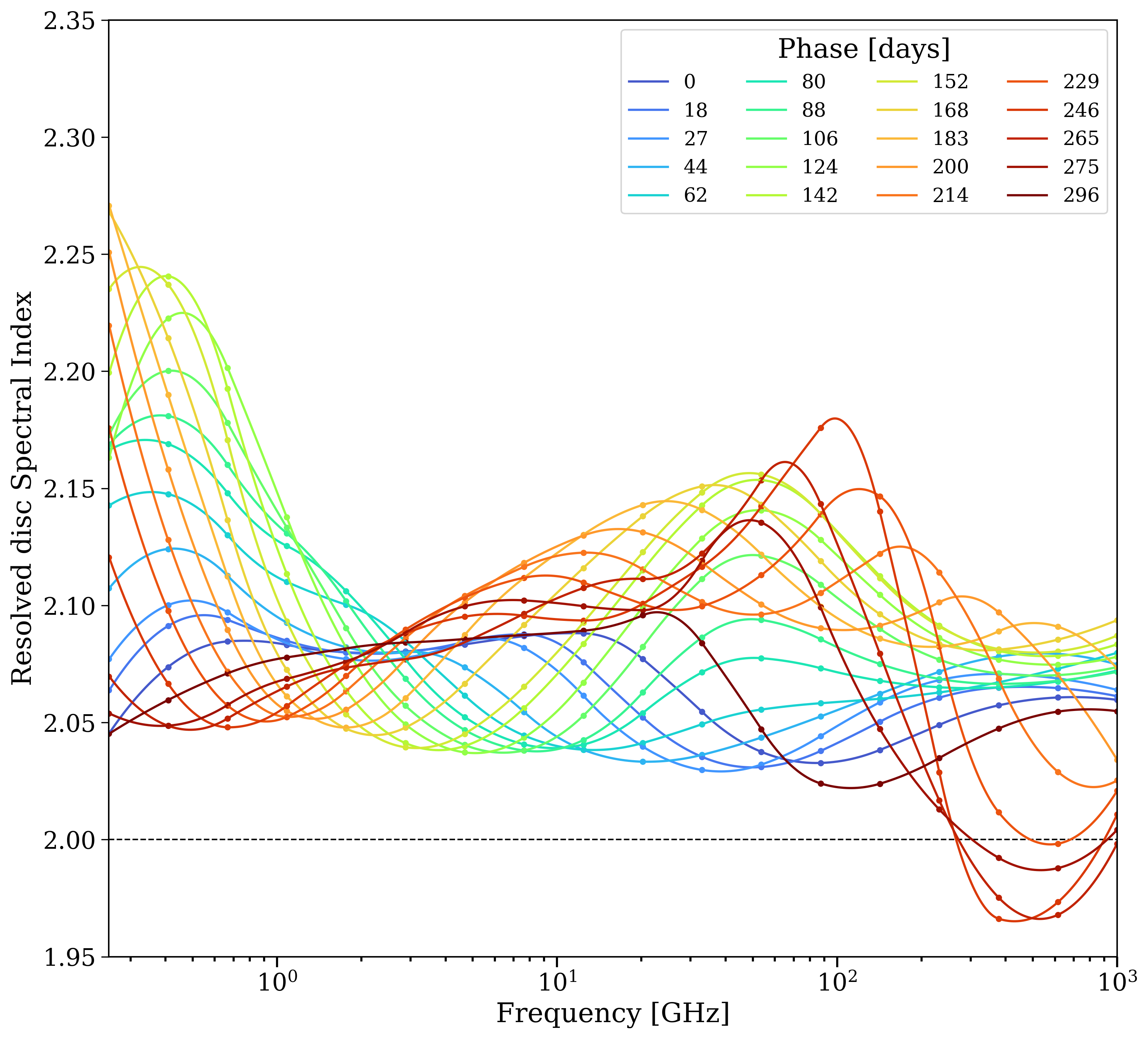}
        \label{fig:resolved spectral index An315u4}
      \end{minipage}

  \vspace{\floatsep} 
  
  \centering
  \begin{minipage}[b]{0.49\textwidth}
    \centering
    \includegraphics[width=\textwidth]{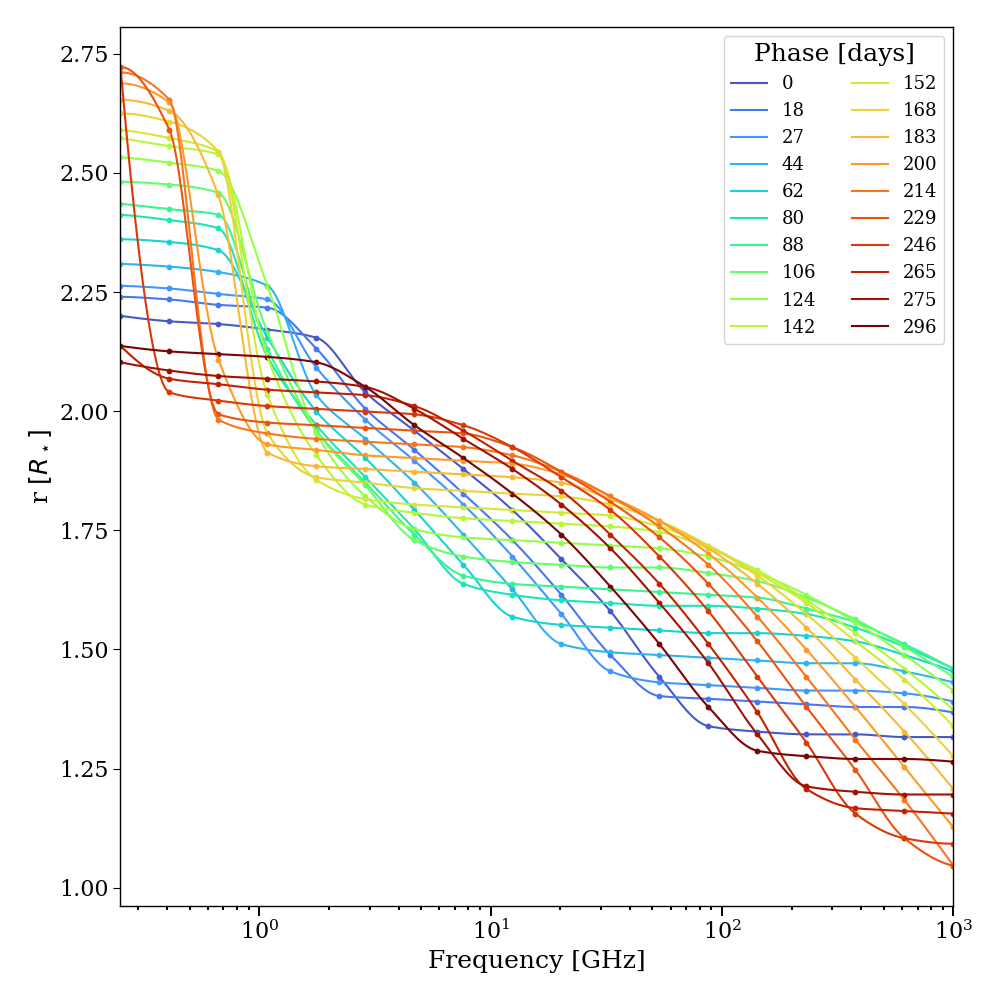}
    \label{fig:radius-freq An315u4}
  \end{minipage}
  \hfill
  \begin{minipage}[b]{0.49\textwidth}
    \centering
    \includegraphics[width=\textwidth]{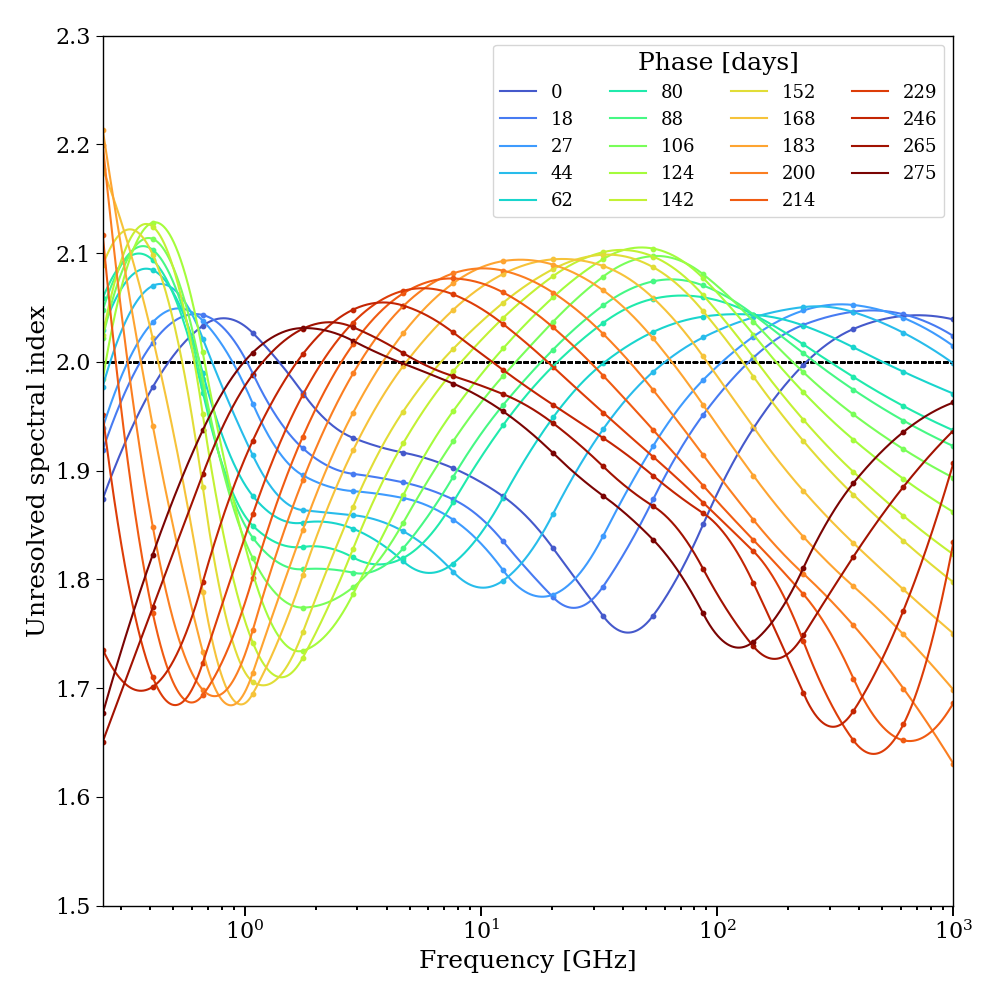}
    \label{fig:unresolved spectral index An315u4}
  \end{minipage}
  
  \caption{Same as Fig \ref{fig:all_plots} for An315u4.}
  \label{fig:all_plots An315u4}
\end{figure*}

\begin{figure*}[htbp] 
  \centering
  \begin{minipage}[b]{0.49\textwidth}
        \centering
        \includegraphics[width=\textwidth]{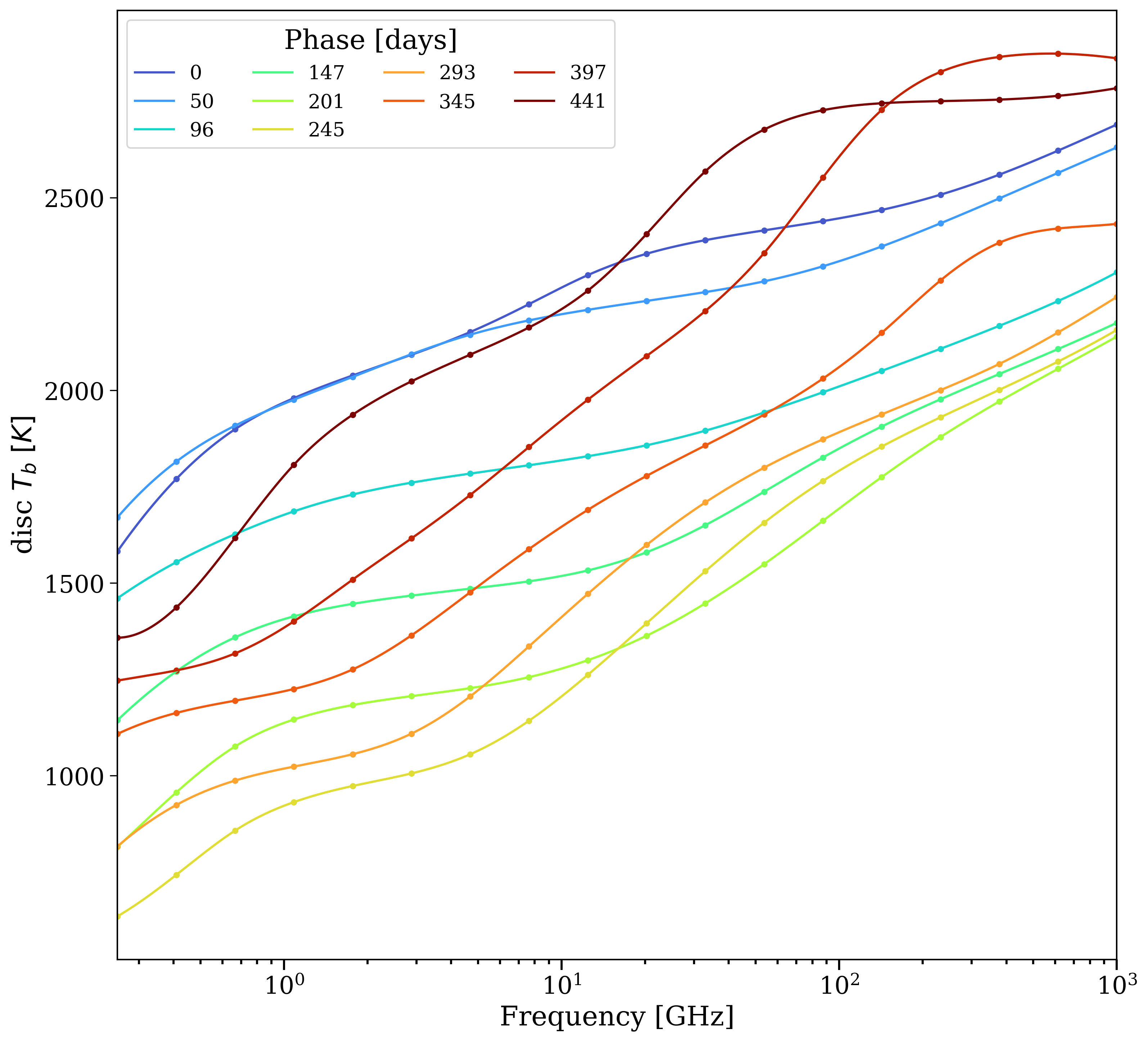}
        \label{fig:disc brightness temperature M2n315u6}
      \end{minipage}
      \hfill
      \begin{minipage}[b]{0.49\textwidth}
        \centering
        \includegraphics[width=\textwidth]{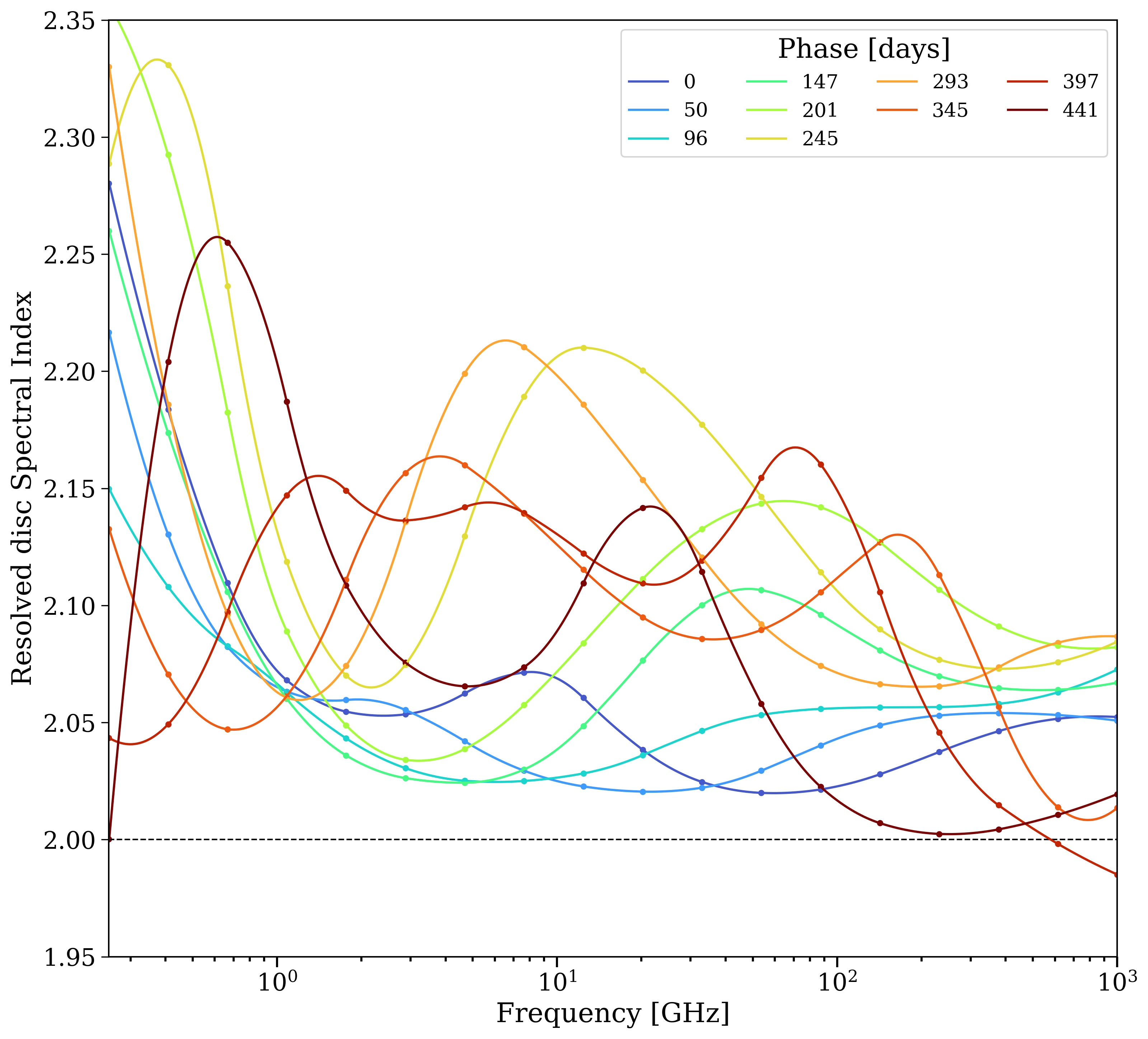}
        \label{fig:resolved spectral index M2n315u6}
      \end{minipage}

  \vspace{\floatsep} 
  
  \centering
  \begin{minipage}[b]{0.49\textwidth}
    \centering
    \includegraphics[width=\textwidth]{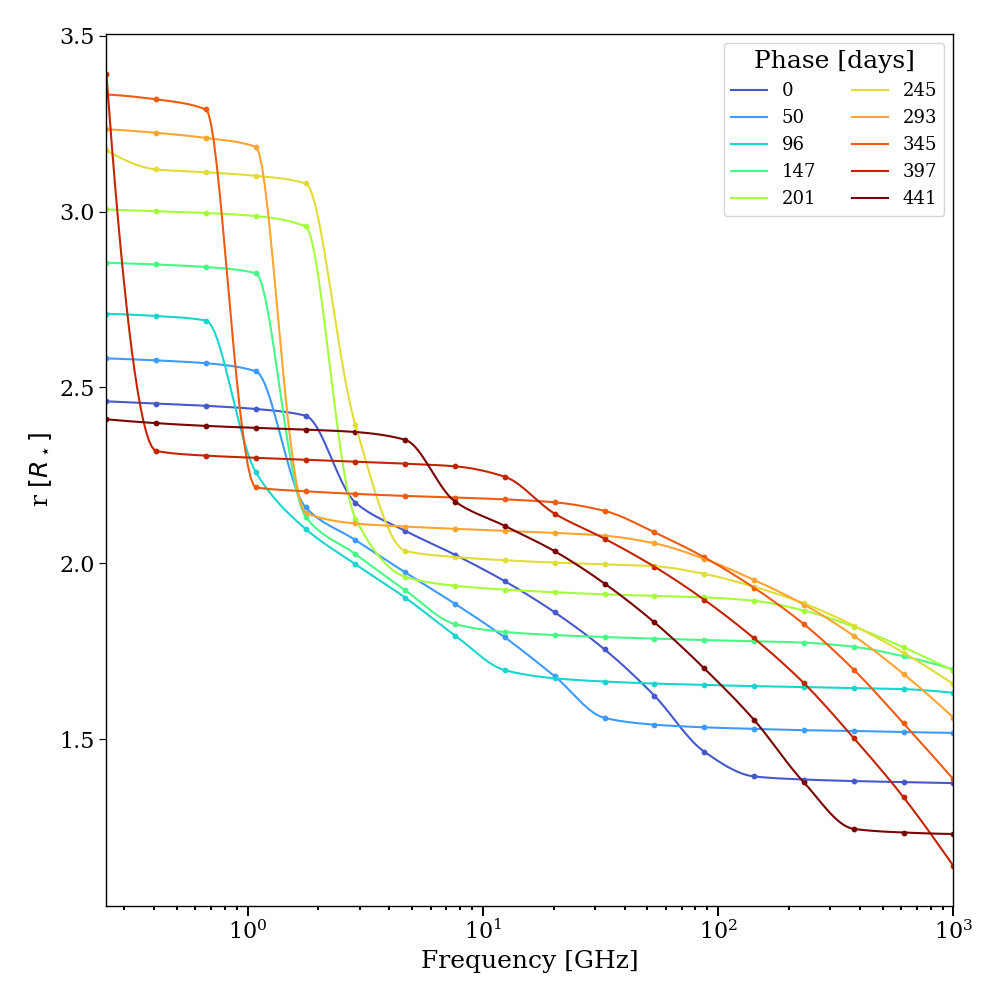}
    \label{fig:radius-freq M2n315u6}
  \end{minipage}
  \hfill
  \begin{minipage}[b]{0.49\textwidth}
    \centering
    \includegraphics[width=\textwidth]{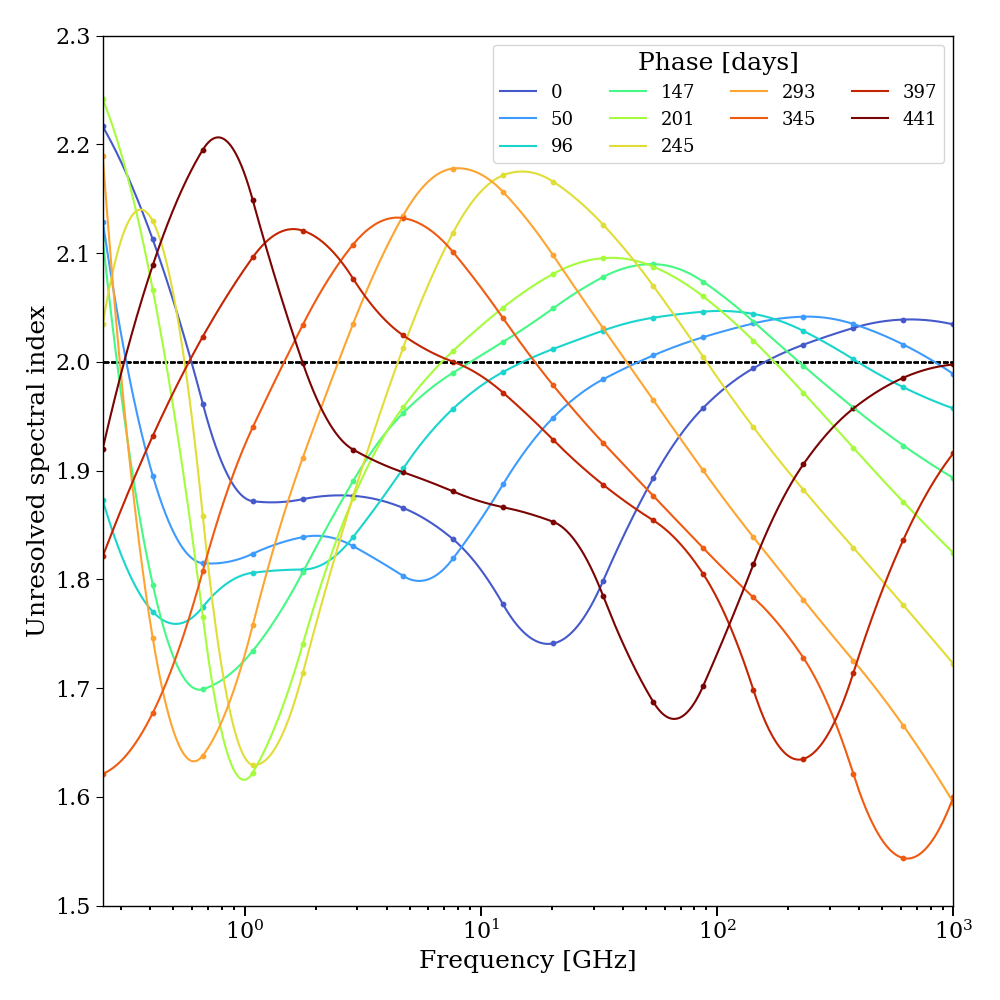}
    \label{fig:unresolved spectral index M2n315u6}
  \end{minipage}
  
  \caption{Same as Fig \ref{fig:all_plots} for M2n315u6.}
  \label{fig:all_plots M2n315u6}
\end{figure*}

\begin{figure*}
    \centering
   \begin{subfigure}
        \centering
        \label{subfig:discTb-T An315u4}
        {\includegraphics[width=0.95\linewidth]{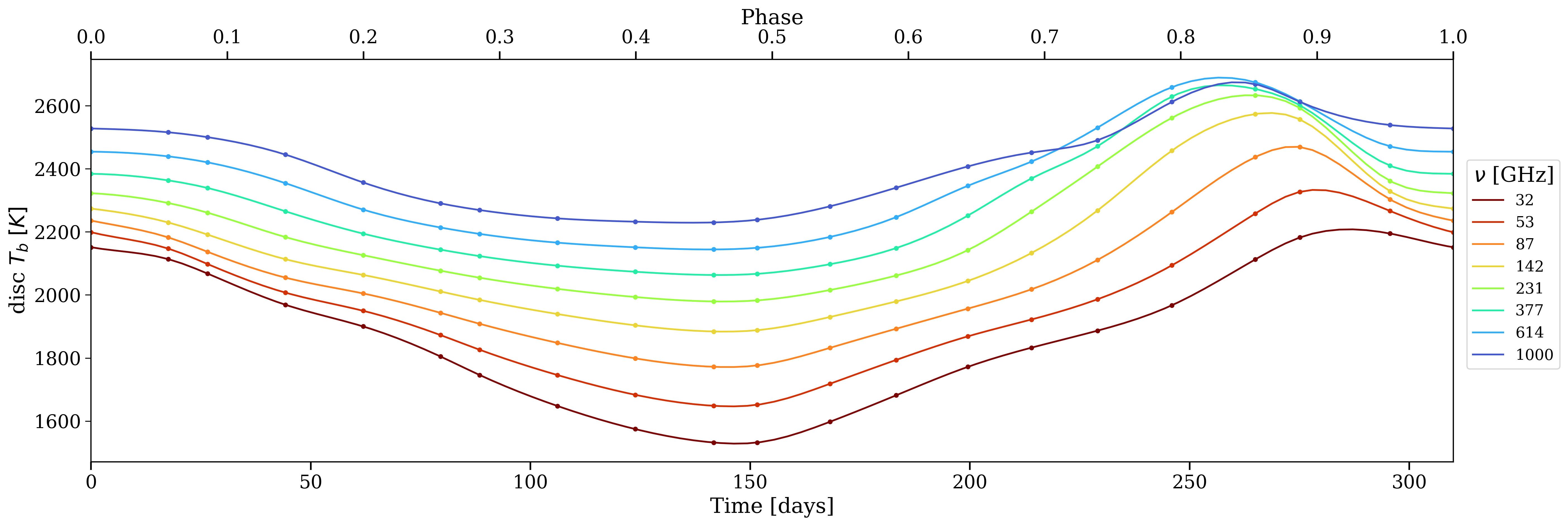}}
   \end{subfigure}
   
   \begin{subfigure}
        \centering
        \label{subfig:radius-T An315u4}
       {\includegraphics[width=0.95\textwidth]{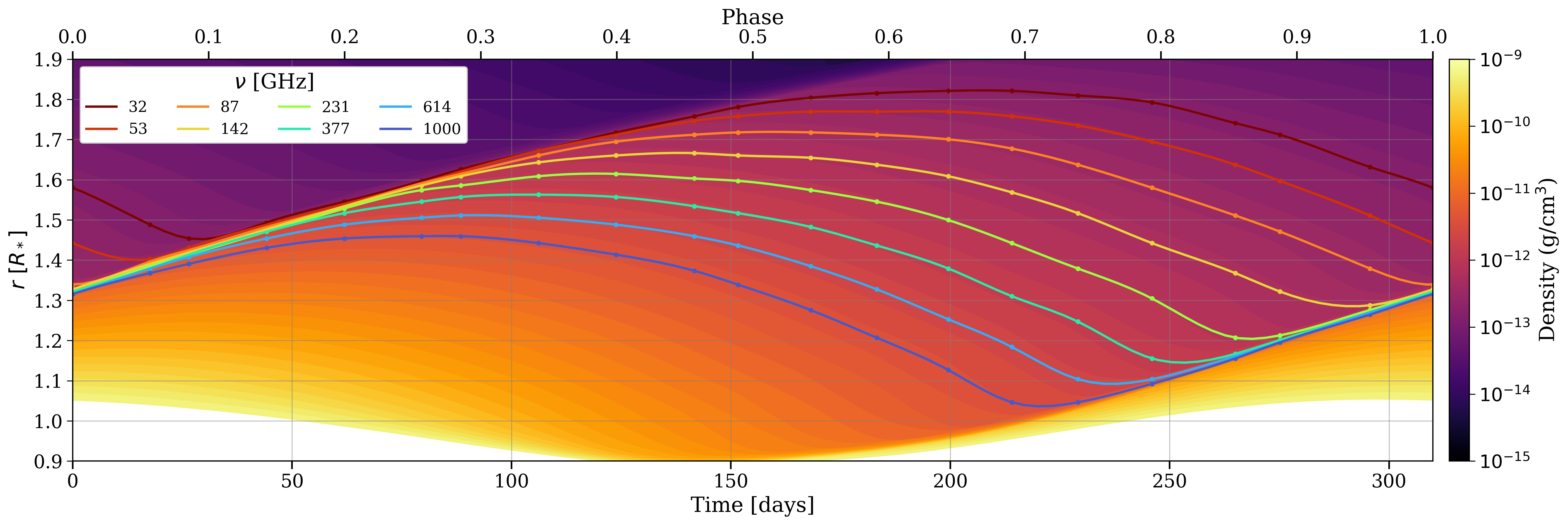}}
   \end{subfigure}
   
   \begin{subfigure}
        \centering
        \label{subfig:FluxDensity-T An315u4}
       {\includegraphics[width=0.95\textwidth]{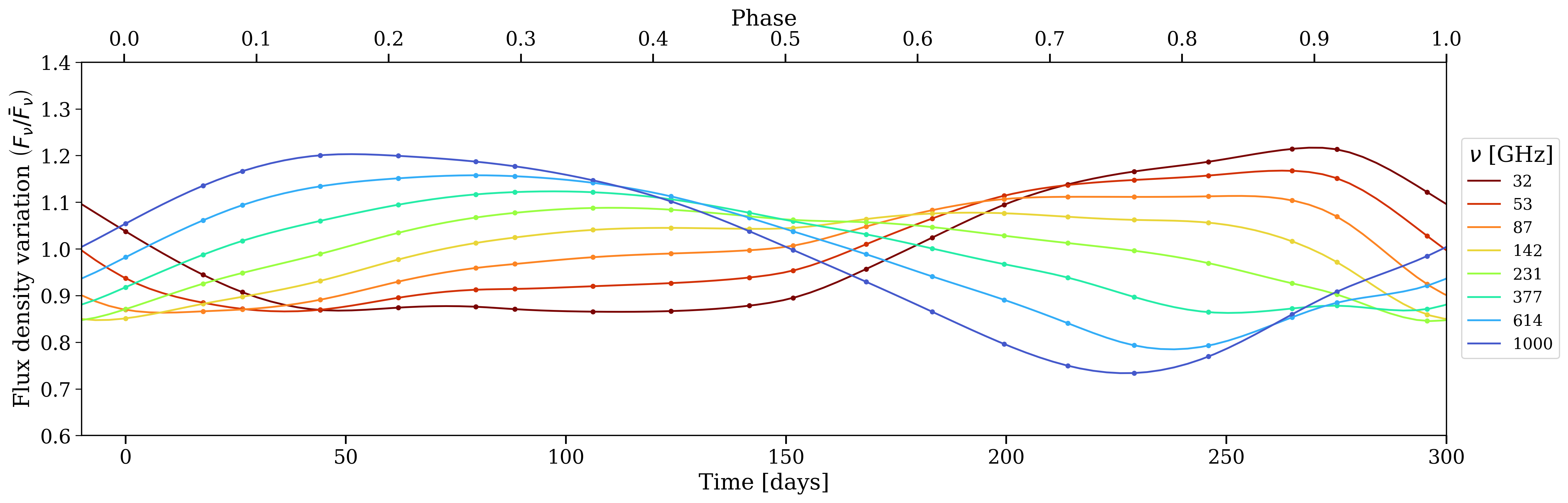}}
   \end{subfigure}

    \caption{Same as Fig.~\ref{fig:temporal_changes} for An315u4.}
\label{fig:temporal_changes An315u4}
\end{figure*}

\begin{figure*}
   \centering
   \begin{subfigure}
        \centering
        \label{subfig:discTb-T M2n315u6}
        {\includegraphics[width=0.95\linewidth]{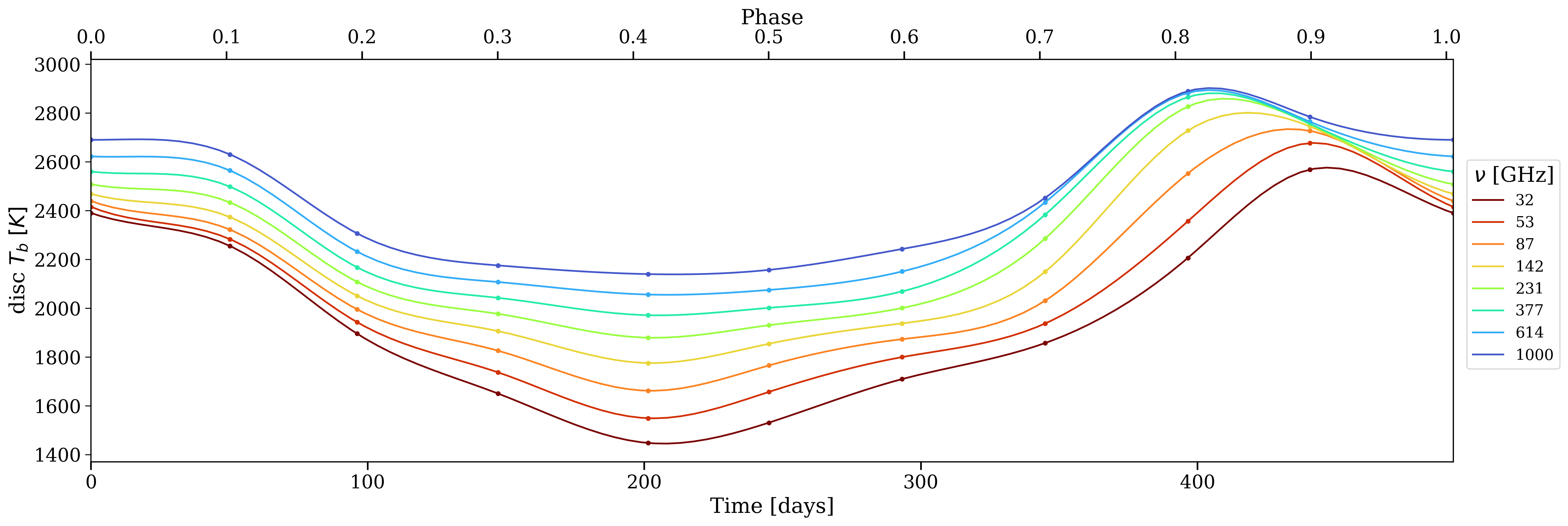}}
   \end{subfigure}
   
   \begin{subfigure}
        \centering
        \label{subfig:radius-T M2n315u6}
       {\includegraphics[width=0.95\textwidth]{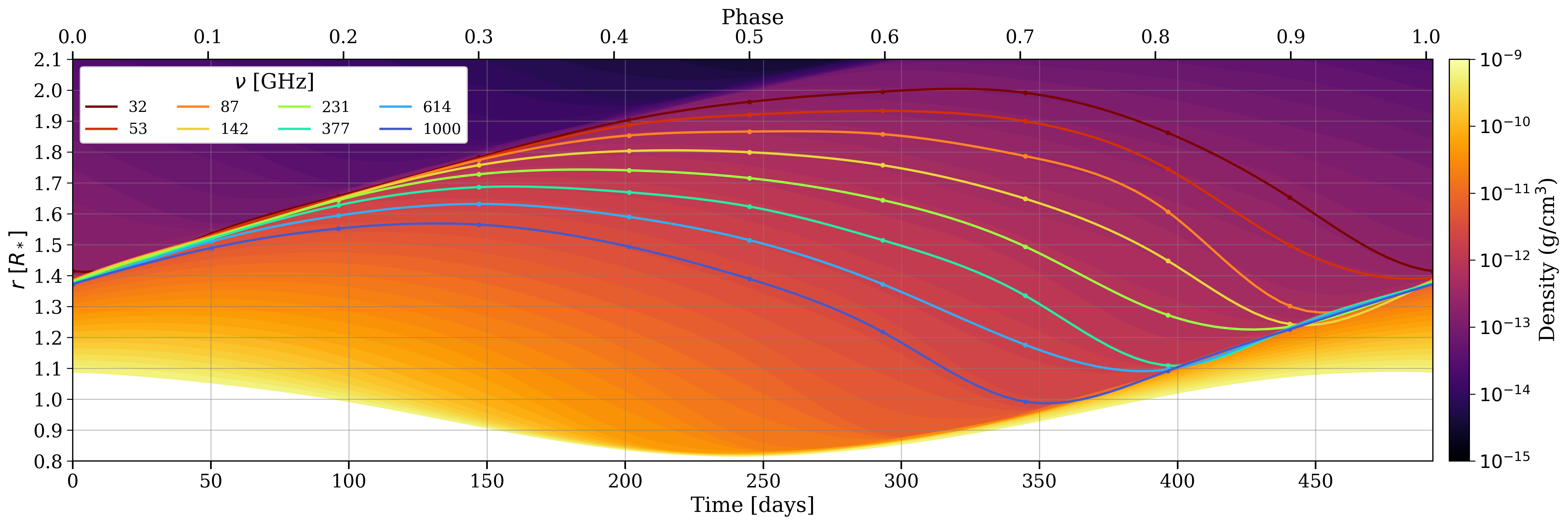}}
   \end{subfigure}
   
   \begin{subfigure}
        \centering
        \label{subfig:FluxDensity-T M2n315u6}
       {\includegraphics[width=0.95\textwidth]{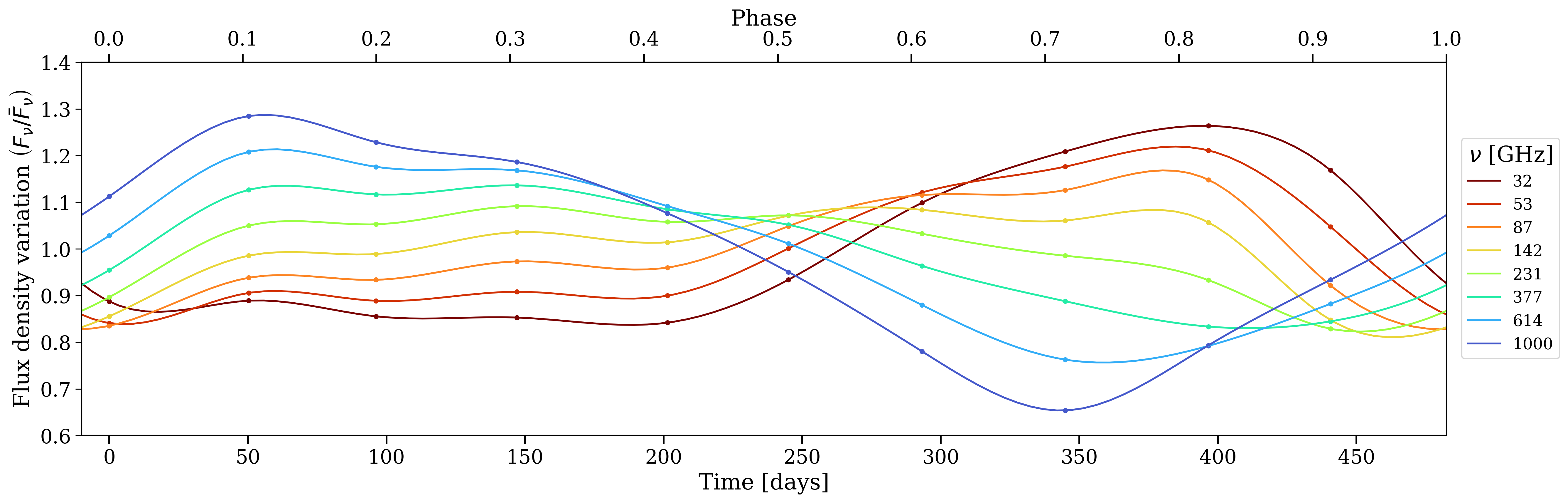}}
   \end{subfigure}

    \caption{Same as Fig.~\ref{fig:temporal_changes} for M2n315u6.}
\label{fig:temporal_changes M2n315u6}
\end{figure*}

Furthermore, we present the result of synthetic observations of models An315u4 and M2n315u6. The temporal variations of radial profiles for gas density and temperature are illustrated in Figs. \ref{fig:An315u4_profile} and \ref{fig:all_plots M2n315u6}, and the brightness temperature profiles and the residuals from the top-hat fits are shown in Figs. 
\ref{fig:profile_residual_An315u4_phases} and \ref{fig:profile_residual_M2n315u6_phases}. The disc brightness temperature, photospheric radius, and resolved and unresolved spectral indices as a function of frequency are depicted in Figs. \ref{fig:all_plots An315u4} and \ref{fig:all_plots M2n315u6}. In Figs. \ref{fig:temporal_changes An315u4} and \ref{fig:temporal_changes M2n315u6}, the observable quantities of disc brightness temperature, photospheric radius, and flux density variations are illustrated as a function of time and phase within one period.

\clearpage 

\begin{figure*}
    \centering
    \includegraphics[width=0.85\linewidth]{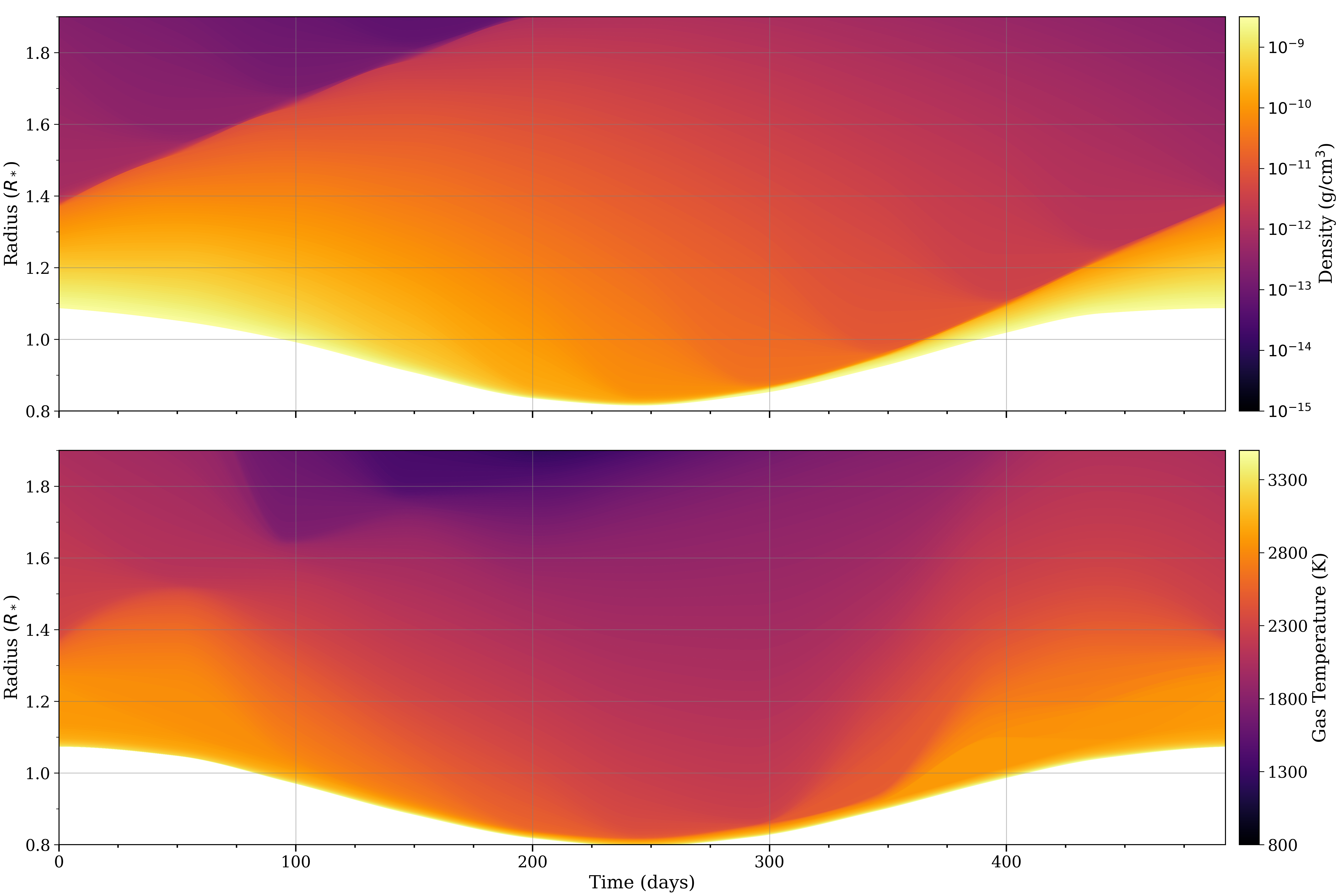}
    \caption{Same as Fig.~\ref{fig:DARWIN_profile} for model M2n315u6.}
    \label{fig:M2n315u6_profile}
\end{figure*}


\begin{figure*}[htbp] 
  \centering
  \begin{minipage}[b]{0.31\linewidth}
    \centering
    \includegraphics[width=1\textwidth]{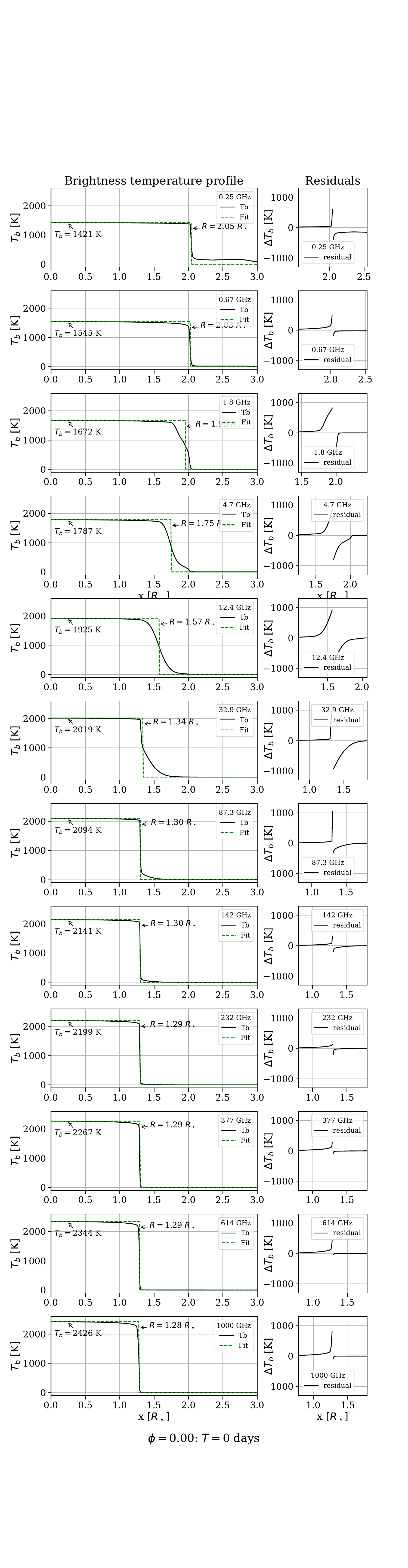}
  \end{minipage}
  \hfill
  \begin{minipage}[b]{0.31\linewidth}
    \centering
    \includegraphics[width=1\textwidth]{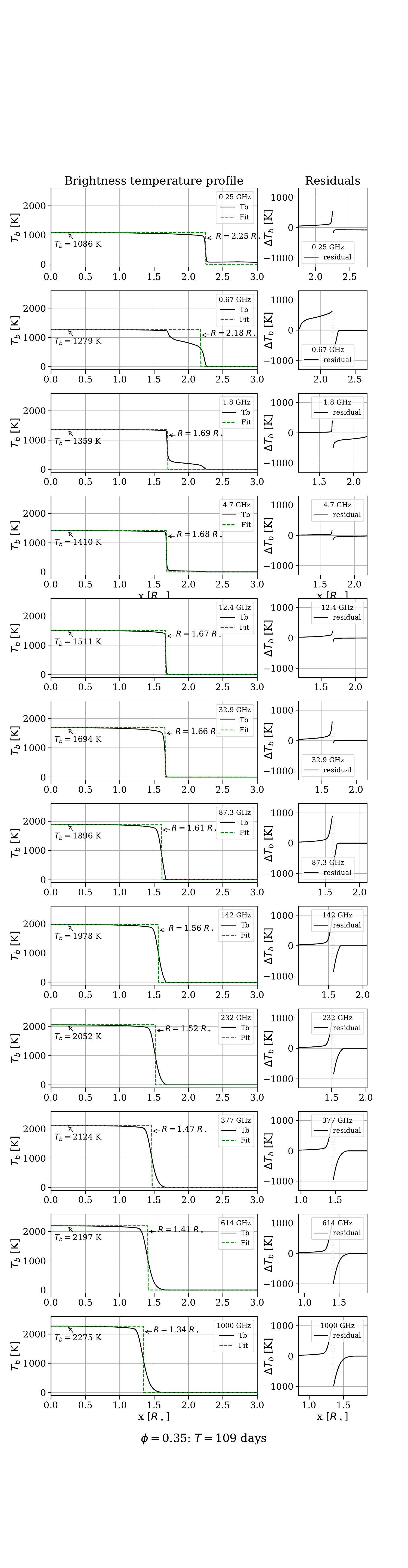}
  \end{minipage}
  \hfill
  \begin{minipage}[b]{0.31\linewidth}
    \centering
    \includegraphics[width=1\textwidth]{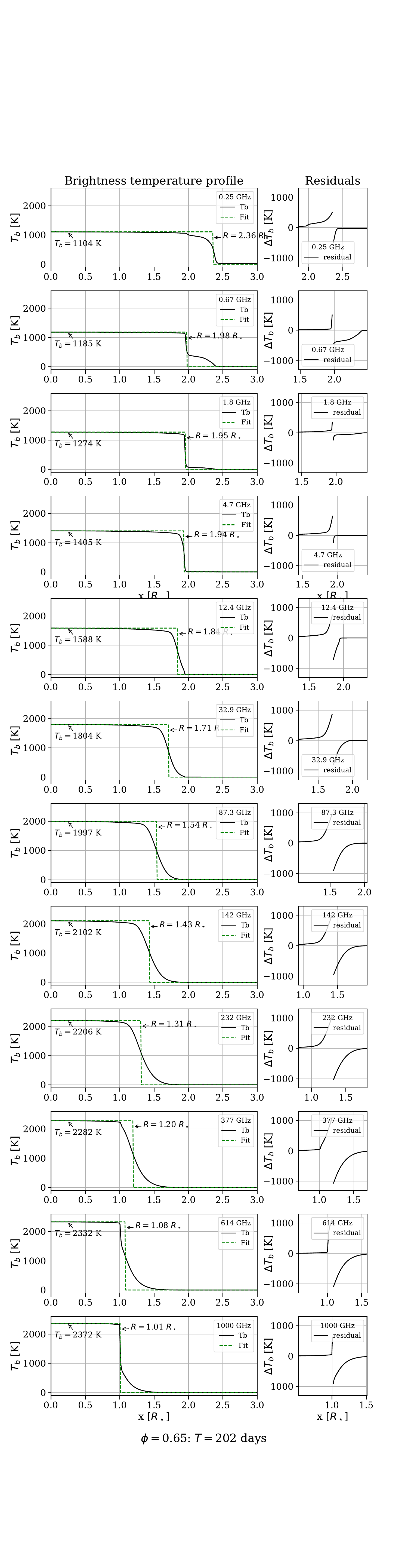}
  \end{minipage}
  \caption{Brightness temperature profiles as a function of radial distance for 12 observation frequencies for model An315u3 at phases 0, 0.35, and 0.65, with annotated fitted radii and disc brightness temperature. The fit residual values are illustrated on the right-hand plots, centered on the resulting radius from fitting.}
  \label{fig:profile_residual_An315u3_phases}
\end{figure*}


\begin{figure*}[htbp] 
  \centering
  \begin{minipage}[b]{0.32\linewidth}
    \centering
    \includegraphics[width=1\textwidth]{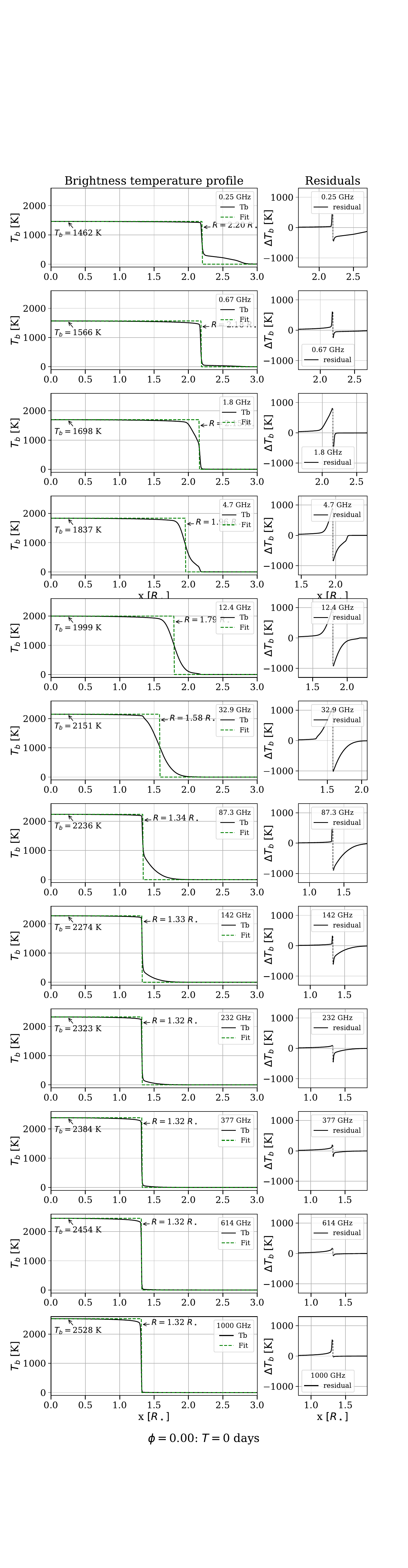}
  \end{minipage}
  \hfill
  \begin{minipage}[b]{0.32\linewidth}
    \centering
    \includegraphics[width=1\textwidth]{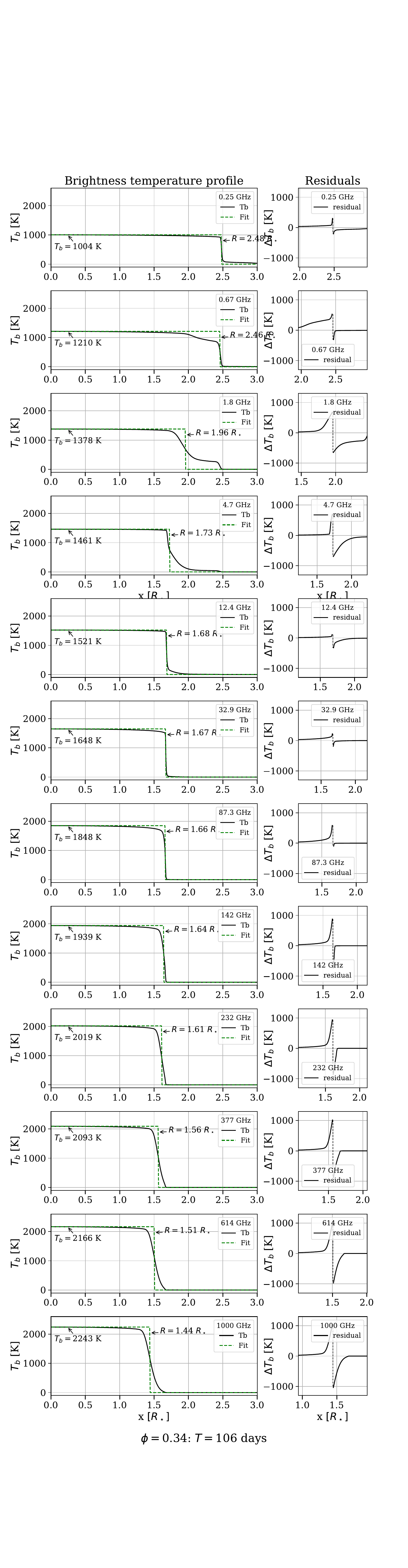}
  \end{minipage}
  \hfill
  \begin{minipage}[b]{0.32\linewidth}
    \centering
    \includegraphics[width=1\textwidth]{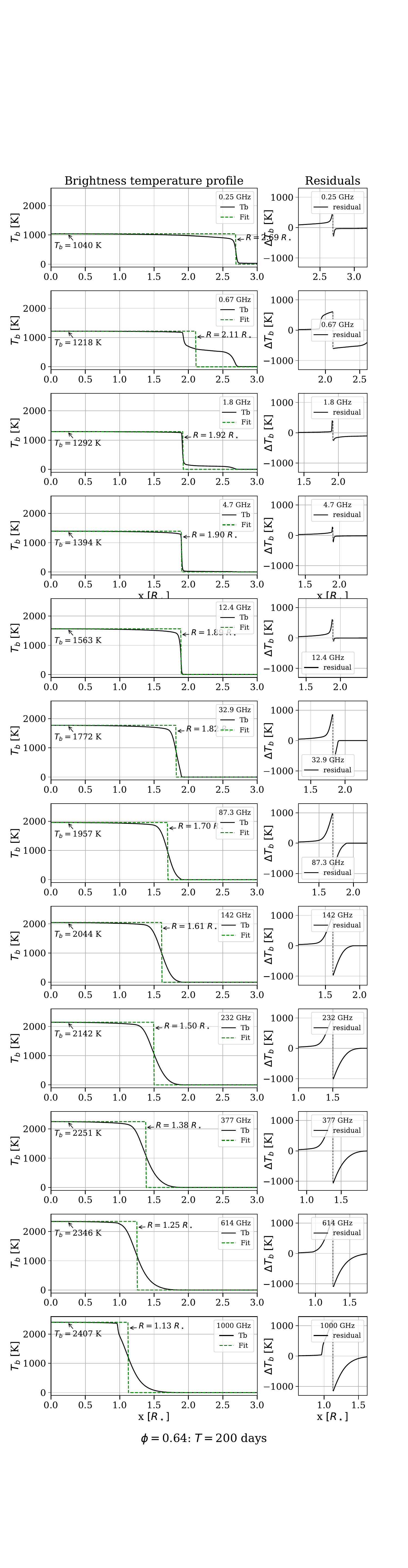}
  \end{minipage}
  \caption{Same as Fig~\ref{fig:profile_residual_An315u3_phases} for model An315u4 and phases 0, 0.34, and 0,65.}
  \label{fig:profile_residual_An315u4_phases}
\end{figure*}


\begin{figure*}[htbp] 
  \centering
  \begin{minipage}[b]{0.32\linewidth}
    \centering
    \includegraphics[width=1\textwidth]{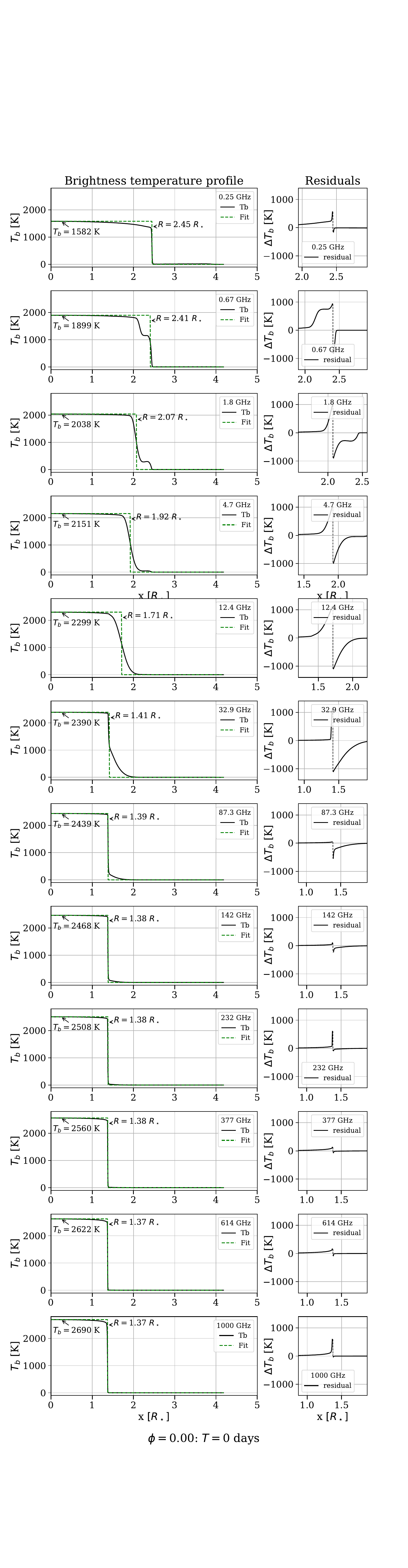}
  \end{minipage}
  \hfill
  \begin{minipage}[b]{0.32\linewidth}
    \centering
    \includegraphics[width=1\textwidth]{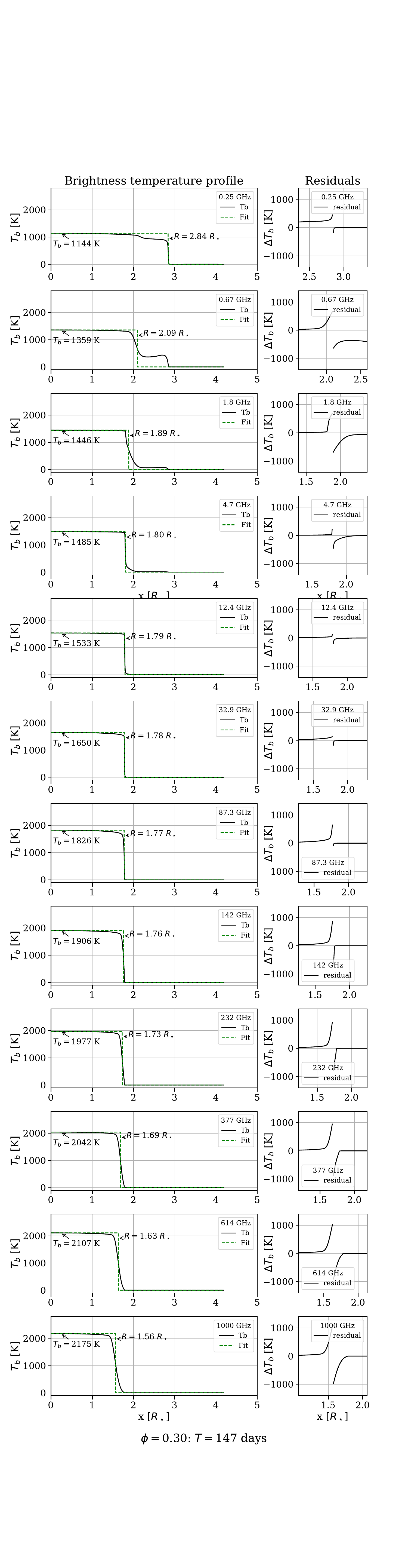}
  \end{minipage}
  \hfill
  \begin{minipage}[b]{0.32\linewidth}
    \centering
    \includegraphics[width=1\textwidth]{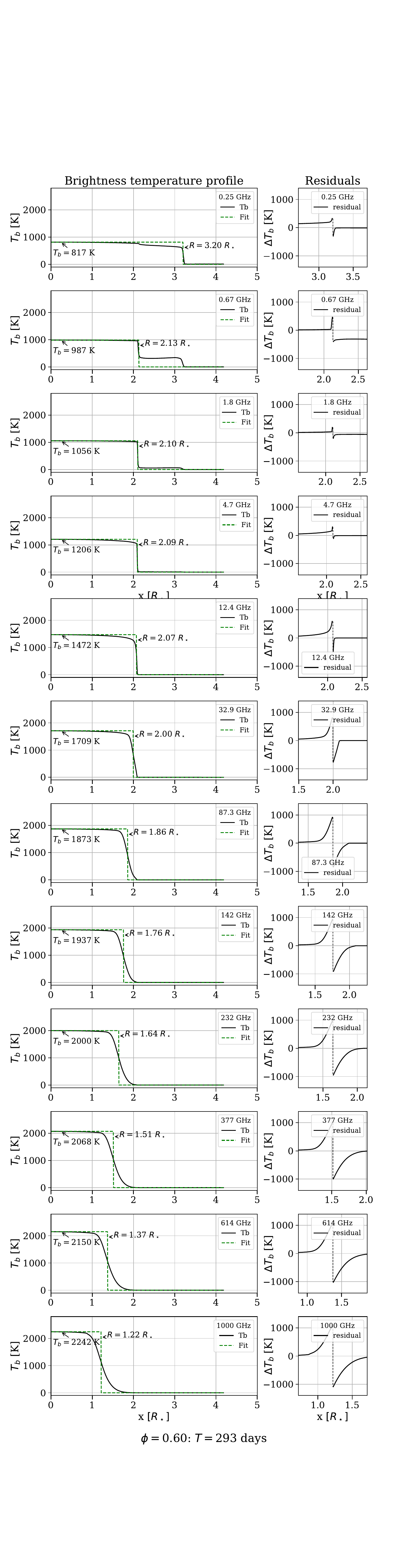}
  \end{minipage}
  \caption{Same as Fig~\ref{fig:profile_residual_An315u3_phases} for model M2n315u6 and phases 0, 0.34, and 0,65.}
  \label{fig:profile_residual_M2n315u6_phases}
\end{figure*}





\clearpage

\clearpage


\bibliography{main}{}

\begin{thebibliography}{}
\expandafter\ifx\csname natexlab\endcsname\relax\def\natexlab#1{#1}\fi
\providecommand{\url}[1]{\href{#1}{#1}}
\providecommand{\dodoi}[1]{doi:~\href{http://doi.org/#1}{\nolinkurl{#1}}}
\providecommand{\doeprint}[1]{\href{http://ascl.net/#1}{\nolinkurl{http://ascl.net/#1}}}
\providecommand{\doarXiv}[1]{\href{https://arxiv.org/abs/#1}{\nolinkurl{https://arxiv.org/abs/#1}}}

\bibitem[{{Braun} {et~al.}(2019){Braun}, {Bonaldi}, {Bourke}, {Keane}, \&
  {Wagg}}]{Braun2019}
{Braun}, R., {Bonaldi}, A., {Bourke}, T., {Keane}, E., \& {Wagg}, J. 2019,
  arXiv e-prints, arXiv:1912.12699, \dodoi{10.48550/arXiv.1912.12699}

\bibitem[{Carilli {et~al.}(2021)Carilli, Mason, Rosero, Butler, Carilli,
  Murphy, \& Walker}]{Carilli2021NextGV}
Carilli, C.~L., Mason, B., Rosero, V., {et~al.} 2021, in Next Generation Very
  Large Array Memos Series.
\newblock \url{https://library.nrao.edu/public/memos/ngvla/NGVLA_92.pdf}

\bibitem[{{Carilli} \& {Rawlings}(2004)}]{Carilli2004}
{Carilli}, C.~L., \& {Rawlings}, S. 2004, \nar, 48, 979,
  \dodoi{10.1016/j.newar.2004.09.001}

\bibitem[{{Dalgarno} \& {Lane}(1966)}]{Dalgarno1966}
{Dalgarno}, A., \& {Lane}, N.~F. 1966, \apj, 145, 623, \dodoi{10.1086/148801}

\bibitem[{{Freytag} \& {H{\"o}fner}(2023)}]{Freytag2023}
{Freytag}, B., \& {H{\"o}fner}, S. 2023, \aap, 669, A155,
  \dodoi{10.1051/0004-6361/202244992}

\bibitem[{{Freytag} {et~al.}(2017){Freytag}, {Liljegren}, \&
  {H{\"o}fner}}]{Freytag2017}
{Freytag}, B., {Liljegren}, S., \& {H{\"o}fner}, S. 2017, \aap, 600, A137,
  \dodoi{10.1051/0004-6361/201629594}

\bibitem[{{H{\"o}fner} {et~al.}(2016){H{\"o}fner}, {Bladh}, {Aringer}, \&
  {Ahuja}}]{Hofner2016}
{H{\"o}fner}, S., {Bladh}, S., {Aringer}, B., \& {Ahuja}, R. 2016, \aap, 594,
  A108, \dodoi{10.1051/0004-6361/201628424}

\bibitem[{{H{\"o}fner} {et~al.}(2022){H{\"o}fner}, {Bladh}, {Aringer}, \&
  {Eriksson}}]{Hofner2022}
{H{\"o}fner}, S., {Bladh}, S., {Aringer}, B., \& {Eriksson}, K. 2022, \aap,
  657, A109, \dodoi{10.1051/0004-6361/202141224}

\bibitem[{{H{\"o}fner} \& {Freytag}(2019)}]{hofner2019}
{H{\"o}fner}, S., \& {Freytag}, B. 2019, \aap, 623, A158,
  \dodoi{10.1051/0004-6361/201834799}

\bibitem[{{H{\"o}fner} \& {Olofsson}(2018)}]{Hofner2018}
{H{\"o}fner}, S., \& {Olofsson}, H. 2018, \aapr, 26, 1,
  \dodoi{10.1007/s00159-017-0106-5}

\bibitem[{{J{\"a}ger} {et~al.}(2003){J{\"a}ger}, {Dorschner}, {Mutschke},
  {Posch}, \& {Henning}}]{Jaeger2003A&A}
{J{\"a}ger}, C., {Dorschner}, J., {Mutschke}, H., {Posch}, T., \& {Henning}, T.
  2003, \aap, 408, 193, \dodoi{10.1051/0004-6361:20030916}

\bibitem[{{Jeong} {et~al.}(2003){Jeong}, {Winters}, {Le Bertre}, \&
  {Sedlmayr}}]{Jeong2003}
{Jeong}, K.~S., {Winters}, J.~M., {Le Bertre}, T., \& {Sedlmayr}, E. 2003,
  \aap, 407, 191, \dodoi{10.1051/0004-6361:20030693}

\bibitem[{{Karakas} {et~al.}(2022){Karakas}, {Cinquegrana}, \&
  {Joyce}}]{Karakas2022}
{Karakas}, A.~I., {Cinquegrana}, G., \& {Joyce}, M. 2022, \mnras, 509, 4430,
  \dodoi{10.1093/mnras/stab3205}

\bibitem[{{Khouri} {et~al.}(2016){Khouri}, {Maercker}, {Waters}, {Vlemmings},
  {Kervella}, {de Koter}, {Ginski}, {De Beck}, {Decin}, {Min}, {Dominik},
  {O'Gorman}, {Schmid}, {Lombaert}, \& {Lagadec}}]{Khouri2016}
{Khouri}, T., {Maercker}, M., {Waters}, L.~B.~F.~M., {et~al.} 2016, \aap, 591,
  A70, \dodoi{10.1051/0004-6361/201628435}

\bibitem[{{Khouri} {et~al.}(2024){Khouri}, {Olofsson}, {Vlemmings}, {Schirmer},
  {Tafoya}, {Maercker}, {De Beck}, {Nyman}, \& {Saberi}}]{Khouri2024A&A}
{Khouri}, T., {Olofsson}, H., {Vlemmings}, W.~H.~T., {et~al.} 2024, \aap, 685,
  A11, \dodoi{10.1051/0004-6361/202348382}

\bibitem[{{Lodders}(2019)}]{Lodders2019}
{Lodders}, K. 2019, arXiv e-prints, arXiv:1912.00844,
  \dodoi{10.48550/arXiv.1912.00844}

\bibitem[{{Marigo} {et~al.}(2020){Marigo}, {Cummings}, {Curtis}, {Kalirai},
  {Chen}, {Tremblay}, {Ramirez-Ruiz}, {Bergeron}, {Bladh}, {Bressan},
  {Girardi}, {Pastorelli}, {Trabucchi}, {Cheng}, {Aringer}, \&
  {Tio}}]{Marigo2020}
{Marigo}, P., {Cummings}, J.~D., {Curtis}, J.~L., {et~al.} 2020, Nature
  Astronomy, 4, 1102, \dodoi{10.1038/s41550-020-1132-1}

\bibitem[{{Matthews} {et~al.}(2015){Matthews}, {Reid}, \&
  {Menten}}]{Matthews2015}
{Matthews}, L.~D., {Reid}, M.~J., \& {Menten}, K.~M. 2015, \apj, 808, 36,
  \dodoi{10.1088/0004-637X/808/1/36}

\bibitem[{Matthews {et~al.}(2018)Matthews, Reid, Menten, \&
  Akiyama}]{Matthews2018}
Matthews, L.~D., Reid, M.~J., Menten, K.~M., \& Akiyama, K. 2018, The
  Astronomical Journal, 156, 15, \dodoi{10.3847/1538-3881/aac491}

\bibitem[{Menten {et~al.}(2012)Menten, Reid, Kami{\'n}ski, \&
  Claussen}]{Menten2012}
Menten, K.~M., Reid, M.~J., Kami{\'n}ski, T., \& Claussen, M.~J. 2012,
  Astronomy and Astrophysics, 543, \dodoi{10.1051/0004-6361/201219422}

\bibitem[{{Murphy} {et~al.}(2018){Murphy}, {Bolatto}, {Chatterjee}, {Casey},
  {Chomiuk}, {Dale}, {de Pater}, {Dickinson}, {Francesco}, {Hallinan},
  {Isella}, {Kohno}, {Kulkarni}, {Lang}, {Lazio}, {Leroy}, {Loinard},
  {Maccarone}, {Matthews}, {Osten}, {Reid}, {Riechers}, {Sakai}, {Walter}, \&
  {Wilner}}]{Murphy2018}
{Murphy}, E.~J., {Bolatto}, A., {Chatterjee}, S., {et~al.} 2018, in
  Astronomical Society of the Pacific Conference Series, Vol. 517, Science with
  a Next Generation Very Large Array, ed. E.~{Murphy}, 3,
  \dodoi{10.48550/arXiv.1810.07524}

\bibitem[{{Reid} \& {Menten}(1997)}]{Reid1997}
{Reid}, M.~J., \& {Menten}, K.~M. 1997, \apj, 476, 327, \dodoi{10.1086/303614}

\bibitem[{{Saha}(1920)}]{Saha1920}
{Saha}, M.~N. 1920, \nat, 105, 232, \dodoi{10.1038/105232b0}

\bibitem[{{Selina} {et~al.}(2018){Selina}, {Murphy}, {McKinnon}, {Beasley},
  {Butler}, {Carilli}, {Clark}, {Durand}, {Erickson}, {Grammer}, {Hiriart},
  {Jackson}, {Kent}, {Mason}, {Morgan}, {Ojeda}, {Rosero}, {Shillue},
  {Sturgis}, \& {Urbain}}]{Selina2018}
{Selina}, R.~J., {Murphy}, E.~J., {McKinnon}, M., {et~al.} 2018, in
  Astronomical Society of the Pacific Conference Series, Vol. 517, Science with
  a Next Generation Very Large Array, ed. E.~{Murphy}, 15,
  \dodoi{10.48550/arXiv.1810.08197}

\bibitem[{Vlemmings {et~al.}(2024)Vlemmings, Khouri, Bojnordi~Arbab, De~Beck,
  \& Maercker}]{Vlemmings2024}
Vlemmings, W., Khouri, T., Bojnordi~Arbab, B., De~Beck, E., \& Maercker, M.
  2024, Nature, 633, 323, \dodoi{10.1038/s41586-024-07836-9}

\bibitem[{Vlemmings {et~al.}(2018)Vlemmings, Khouri, Beck, Olofsson,
  Garc{\'\i}a-Segura, Villaver, Baudry, Humphreys, Maercker, \&
  Ramstedt}]{Vlemmings2018}
Vlemmings, W.~H., Khouri, T., Beck, E.~D., {et~al.} 2018, Astronomy and
  Astrophysics, 613, \dodoi{10.1051/0004-6361/201832929}

\bibitem[{{Vlemmings} {et~al.}(2019){Vlemmings}, {Khouri}, \&
  {Olofsson}}]{Vlemmings2019}
{Vlemmings}, W.~H.~T., {Khouri}, T., \& {Olofsson}, H. 2019, \aap, 626, A81,
  \dodoi{10.1051/0004-6361/201935329}

\bibitem[{{Winters} {et~al.}(2000){Winters}, {Le Bertre}, {Jeong}, {Helling},
  \& {Sedlmayr}}]{Winters2000}
{Winters}, J.~M., {Le Bertre}, T., {Jeong}, K.~S., {Helling}, C., \&
  {Sedlmayr}, E. 2000, \aap, 361, 641

\bibitem[{{Wittkowski} {et~al.}(2017){Wittkowski}, {Hofmann}, {H{\"o}fner}, {Le
  Bouquin}, {Nowotny}, {Paladini}, {Young}, {Berger}, {Brunner}, {de
  Gregorio-Monsalvo}, {Eriksson}, {Hron}, {Humphreys}, {Lindqvist}, {Maercker},
  {Mohamed}, {Olofsson}, {Ramstedt}, \& {Weigelt}}]{Wittkowski2017}
{Wittkowski}, M., {Hofmann}, K.~H., {H{\"o}fner}, S., {et~al.} 2017, \aap, 601,
  A3, \dodoi{10.1051/0004-6361/201630214}

\end{thebibliography}
\bibliographystyle{aasjournal}



\end{document}